\documentstyle[epsfig]{l-aa}
\newcommand {\dx} {{\rm d}x}
\newcommand {\dy} {{\rm d}y}
\newcommand {\hf} {\hat{f}}
\newcommand {\et} {{\it et al.}}

\begin{document} 
\thesaurus{03      % A&A Section 3: Astron. instr., methods and techniques
          (03.13.2;% Methods: data analysis
           03.13.6;% Methods: statistical 
           11.03.1 % Galaxies: clusters: general
          )} 
\title{Density estimation with non--parametric methods.}
\author {Fadda D.~\inst{1}, Slezak E.~\inst{2}, Bijaoui A.~\inst{2}
}
\offprints{Fadda D.}
\institute{
Dipartimento di Astronomia dell'Universit\`a degli Studi di Trieste,\\
SISSA, Via Beirut 4, 34014 Trieste, 
Italy~\thanks{{\it e-mail~:\/}  fadda@sissa.it}
\and
Observatoire de la C\^ote d'Azur, B.P. 4229, F-06304 Nice Cedex 4, 
France~\thanks{{\it e-mails~:\/} slezak@obs-nice.fr, bijaoui@obs-nice.fr}
}
\date{Received date; accepted date} 
\maketitle
\markboth{Fadda \et: Density estimation with 
non--parametric methods}{}

\begin{abstract}

One key issue  in several astrophysical problems  is the evaluation of
the density  probability function underlying an observational discrete
data set. We here review  two non-parametric density  estimators which
recently appeared in the astrophysical literature, namely the adaptive
kernel  density   estimator  and    the  Maximum Penalized  Likelihood
technique, and describe another method based on the wavelet transform.

The efficiency of these estimators is tested by using extensive numerical
simulations  in  the   one-dimensional  case.   The results  are in   good
agreement with theoretical functions  and the three methods appear to
yield consistent estimates.  However, the Maximum Penalized Likelihood
suffers from a lack of resolution and high computational cost due to its
dependency on a minimization  algorithm. The small differences between
kernel and  wavelet estimates are mainly explained  by the  ability of
the  wavelet  method to  take  into  account local   gaps in  the data
distribution.   This new approach   is very promising, since smaller structures
superimposed onto a larger one  are detected only  by this  technique,
especially   when   small samples  are  investigated.    Thus, wavelet
solutions  appear   to be better    suited  for subclustering studies.
Nevertheless,   kernel estimates  seem  more robust  and are  reliable
solutions although some small-scale details can be missed.

In order to  check these estimators  with respect to previous studies,
two galaxy redshift samples, related  to the galaxy cluster A3526  and
to the   {\it Corona Borealis} region,  have   been analyzed.  In both
these cases claims for bimodality are confirmed at a high confidence level.

\end{abstract}

\section{Introduction.}

The galaxy distribution within the local Universe appears to be highly
inhomogeneous.  Groups involving few members, poor or rich clusters
with hundreds of members, and superclusters including dozens of
clusters are common features of the {\it realm of the nebulae}, as are
large regions devoid of bright galaxies (see e.g. Oort 1983, Bahcall
1988, and Geller \& Huchra 1989).  Both kind of structures are defined
as a local enhancement or deficiency of the galaxy density, but the
question of their objective identification is still a matter of
debate.  The amount of subclustering within galaxy clusters and their
present dynamical state is affected by the same problem.  It should be
noted, however, that dark matter today dominates the matter density
according to current theories for galaxy formation.  So meaningful
comparisons between optical data and gravitational distortions
generated by clumps of dark matter would require accurate maps of the
galaxy density. Meanwhile, results coming from gravitational lensing
observations confirm that the distribution of this non--baryonic
component is traced by the galaxy population and the X-ray
emitting intra-cluster gas.  Thus, the matter density also fluctuates
from one location to another.  \\ Besides the estimation of the shape
parameters of galaxy structures, various observational effects on the
galaxy population have been discovered.  The most prominent one is a
morphology-density relation (Dressler 1980) the effects of which are
most noticeable in the high density central regions of galaxy
clusters.  Others effects are discussed with respect to the exact
position of the center of the clusters, i.e. of the peak matter
density.  Beyond these observational trends, astronomers try to
understand the role of environmental effects on the formation and
evolution of galaxies (cf. the origin of cD galaxies).  The local
galaxy density is surely one piece of relevant information for modeling this
environment.  \\

Therefore, it appears that questions such as the proper identification
of galaxy clusters or the  discrimination among different cosmological
scenarios  can be faced providing that  accurate  and reliable density
estimators can be applied  to  galaxy catalogues.    Three-dimensional
studies are still hampered by the lack of redshift data for wide-field
surveys, but  valuable    information  about   the   overall    galaxy
distribution or   the  structure of galaxy   clusters  can be obtained
through surface  densities computed from two-dimensional catalogues or
by  means  of redshift distributions,   respectively.   \\ The various
methods which have been developed in  order to obtain these estimates can
be divided into  two  groups~: parametric methods  and non--parametric
ones.  The former   assume a physical model  controlled by a given
set of parameters which have  to be fitted to the  data (e.g., a  power
law or a Gaussian fit, a King or a de Vaucouleurs profile, etc.).  But
sometimes the  underlying physics is too poorly  known to apply such a
method.  In this case, one  must rely on non--parametric methods,  the
simplest   of which is  an  histogram  calculation.  The main
difference with  respect to  the   previous approach comes   from  the
influence of the  data points $x_i$ on  the estimate at  location $x$.
All the  points have the  same  importance for  parametric estimators,
whereas non--parametric estimators  are asymptotically local, i.e. the
influence of the points  vanishes as the distance from the location where
the density is  computed increases.  Although histograms  fulfill this
condition, these commonly used  estimates present some drawbacks~: the
results change with the bin size and the origin  of the bins.  The use
of better  one-dimensional   density  estimators would  allow  one  to
overcome this kind of problem.  Such estimators already exist but
they are not yet widely used, maybe because the astronomical community
is not aware  of their performances  and limitations.  Therefore, we  plan in
this paper to discuss three of the most promising methods with respect
to one--dimensional applications.   Extensions of the formalism to the
bidimensional  case are straightforward and   are already explained in
the literature (Pisani 1996, Merritt  \& Tremblay 1994, Slezak \et 1993).\\

At least  two of these non--parametric   methods for computing density
estimation have  indeed  been recently described  in  the astronomical
literature.   These   asymptotically local  methods  are the  adaptive
kernel estimator by Pisani (1993) and the Maximum Penalized Likelihood
estimator (hereafter MPL) by  Merritt \& Tremblay (1994).  Another way
to obtain local information  about a  signal is  provided by the wavelet
transform.  Within the astronomical  context,  it is  usually  used to
analyze time series (Goupil  \et 1991,   Norris \et
1994, Szatmary \et 1996) and to  detect structures at various
scales in catalogues (Slezak \et 1993) or images (Slezak \et
1994, Bijaoui \& Ru\'e 1995). Taking advantage of this property,
we   have developed a  wavelet-based  method  in  order  to restore  a
continuous probability density function from a discrete data sample.

Generally,  cluster   analysis  methods  are   sensitive  to different
features  of   the   signal, generating   questions  about   its  real
characteristics.    When  such  a  situation  occurs, a   comprehensive
knowledge of the performances of  each technique is helpful to settle
the discussion.  The recent kernel, MPL and wavelet density estimators
are       based   on    different    sophisticated        mathematical
backgrounds. Whatever the  difficulties in understanding the  related
formulae  in depth may be, detailed tests are  required to get a good insight
into  the validity of   the solutions provided.  Hence, we
decided to compare the results of these three methods by using test cases
of  astronomical   interest.  Knowing conditions   where one algorithm
succeeds better than  the others and the reasons  why it does  so will
allow one to choose the   best estimator for the   considered
data sample.\\

This paper  is organized as follows.  In  the next  section, we briefly
describe the non--parametric density estimators we are testing, namely
the adaptive kernel,  MPL and  wavelet-based  estimators. The
formulae are given  for the  one--dimensional case,  but  most of  the
explanations    stand for   multidimensional analyses   (comprehensive
reviews can be  found  in Silvermann   1986, Scott 1992,  and  Bijaoui
1993). Technical details about the underlying algorithms for computation
of the density estimates are given  in the Appendix.

Then, we
compare their behaviors by using numerical simulations of five different
one-dimensional samples   with  and without noise  (\S~3).  This study
allows us to make  general remarks as  well as detailed comments about
the efficiency of  each method.  These methods  are finally applied to
two real   galaxy redshift  catalogues   in \S~4 and the   results are
discussed with respect  to previous studies.   We give our conclusions
in \S~5.

\section{Non--parametric methods.}

A natural way to get a continuous density function from a discrete set
of data points is  to spread each point  according to a given pattern.
The   linear  smoothing related  to  this    data-based solution is  a
stationary  method,  since the  variations  in  number  density are not
explicitly  taken into account.    Consequently, two kinds of methods
have been  designed to improve  the density estimate.   The first ones
are directly  based on the  data,  since they adapt  the pattern function
on the basis of the local number density.  The second  ones come from
signal processing    theory: the data are   considered  as a function
suffering from a Poisson noise.  A  pioneering example in astronomy of
such an approach is provided by the computation of the distribution of
Cepheid  periods using the  Walsh-Hadamard transform (Bijaoui 1974).\\
The  probability density function  can indeed  be estimated either  by
working on the positions themselves or by analyzing a signal resulting
from  these  positions.  Let  us give  an  example.   Among  available
non--parametric techniques, the \mbox{K-th}  nearest neighbor estimator  was
introduced into astronomy by Dressler  (1980).  If the distances of  $n$
data points  $x_1, x_2, \ldots,  x_n$  to a location  $x$ are ordered,
$d_1(x)\leq d_2(x) \leq \cdots \leq d_n(x)$, this density estimator is
defined by~:
\begin{equation}
\hf_k(x)=\frac{k}{2nd_k(x)}.
\end{equation}
In fact, if the  density is $f(x)$,  one expects to find $2nrf(x)$ data
in  the interval $[x-r,x+r]$ with  $r>0$. By setting $k=2nd_k(x)f(x)$,
we  obtain the definition of  the  estimator. In this way, $\hf(x)$  is
computed at  each  point $x$  with   the same number  of data  points,
leading to  a constant  signal--to--noise   ratio for  the  estimate.  A
similar solution, called  noise cheating image enhancement, was
given by Zweig \et (1975) within the signal processing field.
A  minimum count  value  is first defined,  and the  smallest interval
containing  at least  this count  value  is then determined for  every
location of interest.  The values  of the density are finally obtained
from the summation of the counts divided  by the size of the interval.
Thus, it appears that the difference between the two algorithms lies in the
starting data~; the former deals with raw coordinates, while the latter
processes counts.\\

The \mbox{K-th}  nearest  neighbor density estimate,   as well as   noise
cheating based one, are not perfect: the function is not smooth and it
is not a   probability density since  $\int \hf(x)   \dx$ is infinite.
Hence, they are not appropriate methods  when a global estimate is
required or when one is interested in the  derivatives of the density.
Therefore, better  estimates have been developed  to overcome such drawbacks.
We briefly review in the  following pages three recent and promising methods
which can be used to compute reliable  density estimations.  The first
two  are position-based  methods, while the  third one derives from
signal theory.

\subsection{ Kernel estimators.}

In astrophysical literature the kernel estimator was first used by
Beers \et (1991). An interesting  paper about the estimate of
density profiles by  some non--parametric methods  (including adaptive
kernel and MPL estimators) is that by Merritt \& Tremblay (1994).\\

Let us consider   a  probability  density  function $K(x)$,   i.e.   a
non-negative function normalized to  unity,  and its convolution  with
the         empirical  density function   $f_n(x)=n^{-1}\sum_{i=1}^{n}
\delta(x-x_i)$~:
\begin{equation}
\hf(x)=\int f_n(y) K(x-y) \dy = \frac1n \sum_{i=1,n} K(x-x_i).
\label{eq_kernel}
\end{equation}
This   function is a   kernel estimator  of   the unknown real density
function  $f(x)$ with  $K(x)$ as kernel   function.  We can scale this
estimate by introducing a smoothing parameter $h$, which leads to~:
\begin{equation}
\hf(x)=\frac1n \sum_{i=1}^{n} \frac1h K \left( \frac{x-x_i}{h}
\right).
\label{eq_fixk}
\end{equation}
The estimate $\hf(x)$ is  a probability density function which  shares
the same analytical properties as $K(x)$.\\

The global accuracy of  the estimate $\hf(x)$ can  be evaluated by the
mean integrated square error, defined as~:
\begin{eqnarray}
{\rm MISE}(\hf)&=&E \left[ \int (\hf(x) -f(x))^2 \dx \right] \nonumber \\
&=& \int {\rm Bias}^2(x) \dx + \int {\rm Var}(\hf(x)) \dx.
\end{eqnarray}
It is the   sum  of the  integrated square  bias  and  the  integrated
variance,  the   bias being  the difference between   the true density
$f(x)$ and the estimate $\hf(x)$.    By minimizing this quantity,   an
optimal value for the  parameter  $h$ is obtained.  This value  can be
written as (Silvermann 1986)~: $h_{\rm opt}=c_K\ G(f)$, where $c_K$ is
a constant   depending on the  kernel function  and $G(f)$  a function
related   to the  true  density.  The   best  kernel function with the
constraints $\int K(x) \dx=1$ (normalization)  and $\int x\ K(x)\dx=0$
(symmetry) is the so-called Epanechnickov kernel (1969)~:
\begin{equation}
K_e(x) = \left\{ 
\begin{array}{ll}
\frac34 (1-x^2) & \textrm{if $|x|<1$}\\
0               & \textrm{elsewhere.}
\end{array} 
\right.
\label{eq_epanen}
\end{equation}
Defining    the  efficiency  of   a  kernel    function  as the  ratio
$c_{K_e}/c_K$, one obtains values close to unity for a very large class
of kernels.  So, the choice of the  kernel function must be on the basis
of
other considerations (e.g.,  an high degree of differentiability).  The
choice of  the minimum value of  $G(f)$ involves an assumption on the
form of the true distribution $f(x)$. A usual  choice is the Gaussian
bandwidth that gives the normal reference rule~:
\begin{equation}
h_{\rm opt} \simeq 1.06 \ \hat{\sigma} \ n^{-1/5},
\label{eq_hopt}
\end{equation}
where $\hat{\sigma}$ is the standard deviation of the data.\\

If  we  apply   this   estimator to   data  coming    from long-tailed
distributions,  with a small enough   $h$ to appropriately process the
central part  of  the distribution, a spurious  noise  appears in  the
tails. With  a larger $h$ value for  correctly handling the  tails, we
cannot  see  the details    occurring   in the     main  part of    the
distribution.   In  fact, a   mathematical  derivation shows  that the
integral bias and the integral variance are  proportional to $h^2$ and
to $(nh)^{-1}$, respectively.  Hence,  reducing the  variance produces
an increase  of  the bias, while a  smaller  $h$ reduces the  bias but
enlarges  the  variance.  To overcome   these  defects, adaptive kernel
estimators were introduced. For instance, one can use the estimate~:
\begin{equation}
\hf(x)=\frac{1}{n} \sum_{i=1}^{n} \frac{1}{\lambda_i h} K\left(
\frac{x-x_i}{\lambda_i h} \right),
\label{eq_lambda}
\end{equation}
where $\lambda_i$ are quantities related to the local density at $x_i$
(see the Appendix section for the determination of $h$ and $\lambda_i$
values).  We decided to test this peculiar adaptive kernel estimate.

\subsection{Maximum Penalized Likelihood estimator.}

Applied to the  density  estimation problem, the standard  statistical
technique of Maximum Likelihood proposes to maximize the quantity~:
\begin{equation}
L(g;x_1,x_2,\ldots,x_n)=\prod_{i=1}^{n} g(x_i)
\end{equation}
over the class  of all density functions $g(x)$.  But it fails because
the likelihood can   be  made arbitrarily large  with  density functions
approaching    the empirical density function (i.e.     a sum of delta
functions).\\

An alternative  approach is to penalize the  likelihood by including a
term which describes the roughness of the function, according to the formula:
\begin{equation}
L_{\alpha}(g)=\sum \log g(x_i) - \alpha\ R(g),
\label{MPL_eq}
\end{equation}
where $R(g)$ is a functional and  $\alpha$ is a constant that controls
the  amount  of  smoothing.   Note  that such  a  penalization  of the
likelihood is similar to   the regularization function  introduced for
solving inverse problems  (Titterington   1985, Demoment  1989).   The
estimate $g(x)$ will maximize $L_{\alpha}$  with the constraints $\int
g(x)\dx=1$,   $g(x)\geq 0$ for  every    $x$ and $R(g)<\infty$.   This
approach  makes explicit the two   conflicting  aims in curve  estimation~:
to maximize  the fidelity  to  the data    (the first term  $\sum  \log
g(x_i)$) while  avoiding  rough curves or   rapid  variations, which is
controlled   by the  second term  $R(g)$.    The smaller the  value of
$\alpha$ is, the rougher will be the corresponding MPL estimate.

One can eliminate the necessity for a  positivity constraint on $g$ by
using a penalty  functional based  on the   logarithm of the   density
$f=\log g$. In this  way, $g=\exp (f)$  will automatically be positive.
Moreover, one can assume a penalty functional of the form~:
\begin{equation}
R(g)=\int \left( \frac{{\rm d}^3 \, \log g(x)}{\dx^3} \right)^2 \dx,
\end{equation}
which is equal to zero if  and only if  $g$ is a normal function; in this
way, as $\alpha$ tends to  infinity, the estimate converges towards the
normal  density with the  same mean and variance  as the data.  Hence,
even an overestimate of the smoothing parameter will give, at worst, a
Gaussian fit to the data.  It is possible  to define different penalty
functionals if other kinds of physical functions  are expected for the
problem considered (see, e.g., Merritt \& Tremblay 1994).\\

Once $f= \log g$ is set,  the MPL estimate can  be found by maximizing
the quantity~:
\begin{equation}
\sum f(x_i) - \alpha \int (f''')^2
\end{equation}
with  the constraint   $\int \exp (f(x))   \dx  =1$ (see the  Appendix
 for technical details about the maximization procedure).

\subsection{Wavelets.}

\noindent
In order to derive the formulae related to this approach, which makes use of
the signal theory, let us   consider the convolution of the  empirical
density  with a smoothing function $\phi  (x)$ whose shape and support
will define the resolution of our final estimate~:
\begin{eqnarray} 
\hf_0(x)& = &\int\sum_i\ \frac 1n\ \delta(y-x_i)\ \phi(x-y)\ {\rm d}y
\nonumber \\
&= &\frac 1n \ \sum_i\ \phi(x-x_i).
\label{eq_zero}
\end{eqnarray}
It  appears that $\hf_0(x)$  is identical to  the kernel estimate (see
eq.~\ref{eq_kernel}) providing  that the kernel function is $\phi(x)$.
But the main    difference from the previous   approach  is that  the
positions $x_i$   are only used to  compute   $\hf_0(x)$ on a discrete
grid.  Let us indeed map the interval on which the function is defined
to   $[1,m]$ and  consider  the   values of $\hf_0(x)$   on   the grid
$1,2,\ldots,m$.\\

\noindent
Our  discrete  signal  $\hf_0(k)$   can be locally   analyzed   from a
multi-scale point  of view by using  the wavelet transform.    Within the
peculiar multi-resolution formalism  developed  by  Mallat (1989),  the
signal is  viewed as a set  of details of different sizes superimposed
onto   a very smooth  approximation  at  the   largest scale.  Such  a
space-scale modeling relies on the decomposition of $\hf_0(k)$ on a
set of basis functions for each scale $a_i>$~1 under scrutiny~:
\begin{equation}
 \hf_{a_i}(k)= \langle\ \hf_0(x),\phi_{a_i}(x-k)\ \rangle,
\end{equation}
each basis corresponding to the translations  of dilated versions of a
unique scaling function $\phi(x)$~:
\begin{equation}
\phi_{a_i}(x)=\frac{1}{a_i}\ \phi\left( \frac{x}{a_i} \right).
\end{equation}
The  meaning of the  wavelet coefficients $W_{a_i}$  at scale $a_i$ is
then straightforward~: at each  location they  measure the information
which vanishes  between the approximation at scale  $a_i$ and the next
coarser one  at  scale $a_{i+1}$.   Hence,  these coefficients can  be
easily computed by stating that~:
\begin{equation} 
W_{a_i}(k) = \hf_{a_i}(k) - \hf_{a_{i+1}}(k),
\label{coeff_eq}
\end{equation}
and   consequently the  initial function can    be restored by  a mere
addition   of   these wavelet  coefficients   and   of  the  smoothest
approximation obtained.\\

However, our problem is to recover the probability density function of
the underlying  unknown distribution $f(x)$ from   a limited number of
observational  data points  $\{x_1,x_2,..., x_N\}$.   Lacking external
information,  a  strict  data analysis constrains    one  to take into
account  the  Poisson  noise  these  data   are  suffering  from, while
searching for the best solution consistent with  the data set.  Within
the  vision model related  to  the wavelet approach,   one has thus to
check at each scale whether the enhanced  details are significant with
respect to chance clustering of points. One strategy is provided by the
computation of  the distribution  of the  wavelet coefficients  for  a
locally   uniform  density~: only coefficient   values   with a chance
probability  lower than the value  chosen for  the detection threshold
are to  be  considered as  related to a  genuine  signal (see Bijaoui \&
Ru\'e  1995 and  references    therein).   From the set  of    wavelet
coefficients   $\mathcal{W}$,  a   set  of  thresholded   coefficients
$\mathcal{W}\mathnormal{_t}$   can be obtained  by   rejecting all the
coefficients which  are   not significant.  Below, this
procedure will be  denoted  by the  projection  operator  $P$ such  as
$\mathcal{W}\mathnormal{_t=P[}\mathcal{W}]$.\\

It should be noted that this thresholding strategy is different from a
data compression approach relying on the energy content of the wavelet
coefficients (e.g.  Donoho \et  1993, Pinheiro \& Vidakovic
1995, Vannucci
1996)~\footnote{These~papers~are~retrievable~at~the~web~address~:~http://
schinkel.rz.uni-potsdam.de/u/mathe/numerik/links/wavelets
.1.8.95.html}.  In our opinion, the statistical significance of the
coefficients must indeed be computed locally according to the mean
density at the examined scale and location, and not with respect to
the variance of the coefficients squared (energy content) at this
scale.  In fact, low wavelet coefficients may be locally meaningful
although they represent only a small percentage of the global energy,
and rejecting them will affect the accuracy of the final estimation.
So our solution makes use of the whole set of scales without any
assumption about the regularity of the function sought for, whereas the
smallest scales are explicitly discarded in Pinheiro \& Vidakovic
(1995) (see also \S~4).\\

The key issue is obviously the computation  of the values expected for
wavelet coefficients corresponding to a Poisson  process.  One can try
to perform  Monte-Carlo simulations, but we  preferred to take advantage
of  the Anscombe formula  (Anscombe 1948) which  enables one  to obtain a
distribution with     a  nearly  constant variance    from   a Poisson
distribution   with a large  enough   mean  (above  $\sim$ 10)~;   the
transform of  a function $F(k)$ is  defined  as~: $F_{\rm  A}(k) = 2 \
\sqrt{F(k)+3/8}$.  In order to avoid  error propagation, we decided to
apply this transform  to each successive approximation $\hf_{a_i}(k)$
involved in the ``\`a trous'' algorithm (see Appendix), rather than to
modify only the first approximation $\hf_0(k)$  and run the algorithm
in a  straightforward    way.   This  results   in   modified  wavelet
coefficients which have  the same variance  at each location,  so that
significant values can now be identified by using a classical k--$\sigma$
thresholding. Owing to the  linearity  of  the wavelet transform,   the
threshold at  each  scale can  be computed from  the  variance  of the
wavelet coefficient values at the first scale, where noise dominates.
This  variance can  be computed  either  from the experimental data or
from the theory (Starck \& Bijaoui  1994). These advantages led us
to define the projection operator $P$ in  this regularized space. 
But the values of the wavelet coefficients themselves have still to be taken
from the usual wavelet space, since the non-linearity of the Anscombe
transform will otherwise prevent the use of the restoration algorithm
which is sketched out below and described in more detail in the Appendix.\\
The values of the wavelet coefficients are correlated inside regions whose size
increases with the scale $a$ when no decimation occurs, so that the
result of any statistics involving joint distributions will be incorrect if
this correlation is not properly taken into account.
But such a correlation does not affect the confidence level of a single
wavelet coefficient.
Now each detected structure can be characterized by the confidence level
attached to the single peak value of the wavelet coefficients inside the
connex domain defining its spatial extent.
Moreover, the distance between these extrema values is large enough to
insure their mutual independence.
Hence, as discussed in Slezak \et (1993) and supported by numerical tests,
the statistical significance of the detected structures leading to our
probability density estimate is pretty correct when computed from the
variance of the wavelet coefficients.\\
One must also notice that
the real   statistics of the noise  are  not Poissonian,  but follow a
Bernouilli process, since the relevant information we are interested in
is not the   total number of events,  which  is already known   in the
numerical simulations,  but the distribution  itself, i.e. the density
probability  function. If we consider  a  Poissonian noise, the standard
deviation  of the wavelet coefficients is  overestimated at very large
scales.  But our results  are not biased by  such an effect since we have
considered   only  scales   smaller  than   those  affected   by  this
overestimation.\\

\noindent
In dealing       with        thresholded       wavelet       coefficients
$\mathcal{W}\mathnormal{_t}$,      the  usual straightforward  inverse
wavelet   transform,  which adds  all    the  details  to the  coarsest
approximation, is no longer the way to arrive at an  exact solution.  One has
to look  for a regularization  method which  insures that  the wavelet
transform of the solution  again gives the observed coefficients inside
the significant regions.  Several methods  are available, depending  on
the quantity  which is minimized.   We chose   to apply the  conjugate
gradient technique,  which  looks for the   solution  with  minimal
energy for   the    difference  between the   initial    and  restored
coefficients.   A full description  of  the related  algorithm can  be
found in Ru\'e \& Bijaoui  (1997)~; its  main  steps are given in  the
Appendix.\\ It should first of all  be noted  that the final  density
estimate  is  obtained  at each location   $k$  from selected  wavelet
coefficients  at different scales.  Thus,  several  scales are kept for
 computing our wavelet-based estimate, whereas  only one scale is
used  at  each location $x_i$    with the adaptive  kernel  technique.
Secondly,  the local value of  the adaptive kernel estimate comes from
the   sum  of  kernels  located at   data  points $x_i$,   and a local
underdensity  can never be  explicitly related to  a set of kernels;
such voids   are   only  defined with  respect   to   the neighboring
overdensities.    On the contrary,  negative  wavelet coefficients are
generated by local   voids in the  data  set.  Hence, the  probability
density  function  can be described  by the  wavelet-based approach as
composed of  over-  and  underdensities.   When  these underdensities  are
important features   of the  signal, such  a  capability is  surely an
advantage.

\section{Numerical simulations.}

Every density estimator    has two conflicting  aims~: to  maximize the
fidelity to the data and to avoid roughness or rapid variations in the
estimated curve.     The smoothing parameter  $\lambda$,   the penalty
parameter $\alpha$ and  the threshold parameter  $k$ control these two
aspects  of  the estimation  in kernel,  MPL  and  wavelet estimators,
respectively.  The choice  of these parameters  has to  be made in  an
objective way,  i.e. without using {\it  a priori} knowledge about the
distribution  considered. This  is    possible  by using    data-based
algorithms, e.g. the unbiased cross validation  for the kernel and MPL
estimators  or  the k-$\sigma$ clipping  for  the wavelet coefficients.
These three   estimators favor Gaussian   or quasi-Gaussian
estimates    because of the use  of   a quasi-Gaussian  kernel, of the
adopted form of the penalty functional, and of the shape of the chosen
scaling function.

As for the practical  development of the codes,  we have chosen for the
kernel   estimator     an   Epanechnikov   kernel     function   (see
eq.~\ref{eq_epanen}), which  offers computational advantages because of
its compact support.

In the case of the  wavelet estimator, we  have treated the borders of
the interval with a mirror of the data and we have chosen an  initial grid of 1024
points  in  order  to  recover the  finest   details  of the  examples
considered.   Thus, our results   are not hampered  by any  artificial
smoothing related to information lacking at small scales.  We have also
decided  to threshold the   wavelet  coefficients by using  a  level  of
significance of 3.5 standard deviations (i.e., the probability of getting a
wavelet coefficient  $W$ greater than  the observed value is less than
$10^{-4}$).

In the case of the MPL, the solution greatly depends on the algorithm
of minimization  used. We have obtained  good results with the routine
NAG  E04JAF (see  also   Merritt \& Tremblay   1994).  Obviously,  the
computational time  increases with the number of   points of the curve
which are considered,  i.e.  with  the number  of  parameters  of  the
function to   be   minimized.  A good  compromise   between  resulting
resolution and required  computational time is  to use  50
to  100  points. Though  the MPL  method  is  very attractive from  a
philosophical point of view, its practical usage is penalized by these
difficulties in  minimization. In fact,  an extension of the method
to a two-dimensional case would  become a very hard computational task
on account of the high number of parameters involved.

\subsection{ Description of the samples.}

We  decided  to test   the  previously described  density estimators  by
performing some  numerical simulations  on several  density functions.
We considered five examples,  covering  a large range of  astrophysical
problems~:\\ 
A.~--~a  Gaussian  distribution~: $N(0,1)$~;\\  
B.~--~two similar Gaussians~: ($0.5\  N(0,1)+0.5\ N(3,1)$)~;\\ 
C.~--~a  Gaussian with a small Gaussian in  the tail~: ($0.9\ N(0,1)
+0.1\ N(3,0.5)$)~;\\
D.~--~a  Gaussian  with a    narrow Gaussian  near  its  mean~: ($0.9\
N(0,1)+0.1\  N(1.5,0.1)$)~;\\ 
E.~--~a uniform distribution featuring a Gaussian hole~: $f \propto 
(1-5\sqrt{2\pi}\ N(0,1)/6)$.\\ 
The notation $N(\mu,\sigma)$ stands for a normal random deviate with
a mean of $\mu$ and a standard deviation of $\sigma$.
One can   find  these  distributions    by analyzing   the    velocity
distributions of  galaxies in galaxy clusters  or of stars in globular
clusters.   In   particular, two similar    Gaussians  may represent a
merging between two clusters, while a Gaussian  with another small one
may be found in subclustering cases.  Finally, the hole may correspond
to a local void in a galaxy distribution.

The estimators have to restore  a smooth density function from limited
sets of data points, so that the estimate suffers from noise depending
on the size of the sample.  Moreover, the simulations are generated by
the usual random  routines,  which may sometimes lead  to experimental
data sets in strong   disagreement with the  theoretical distribution.
Therefore,  the quality   of  the restored  function  must be  checked
against the number of data points involved (accuracy) and the fidelity
of the sample  (robustness).  One way  to get a  perfect sample for  a
number  $N$ of events is to  consider the canonical transform $X=F(x)$
where $F(x)$ stands for the repartition function. The $[0,1]$ interval
is divided into $N+1$ equal intervals, which yields a set of $N$ nodes
$x_n$  by  using   the  inverse   transform.    At   these  nodes, the
Kolmogorov-Smirnov test is  satisfied~: by construction,  the distance
between the repartition function  and the cumulative function is equal
to zero.  Hereafter such samples are called ``noiseless'' samples.  In
order to  take into account  the noise coming  from the finite size of
the  samples, we considered three data  sets with an increasing number
of points.  In the pure Gaussian example we  chose a minimum number of
30 points, below  which we decided that the  restoration of the parent
distribution is too difficult, and two more complete sets with 100 and
200 points,  respectively.  We considered  50, 100, and  200 points in
the second   and third  examples, whilst  in    the fourth  example we
consider 100, 200, and 400 points in order to get a high enough signal
for detecting the small feature.  Finally, in the case of the hole, we
 considered a  uniform distribution  and  discarded  the 50,
100, and 200   points  which  fell in   the  region of   the  Gaussian
hole. Hence, the number of points is on  average 430, 860, and 1715 in
the three cases. The width of  the interval was doubled
in order  to avoid having edge effects in  the central  part coming from 
discontinuities at the limits of the uniform distribution.

\begin{figure*}
%\picplace {22cm}
\centerline{\psfig{file=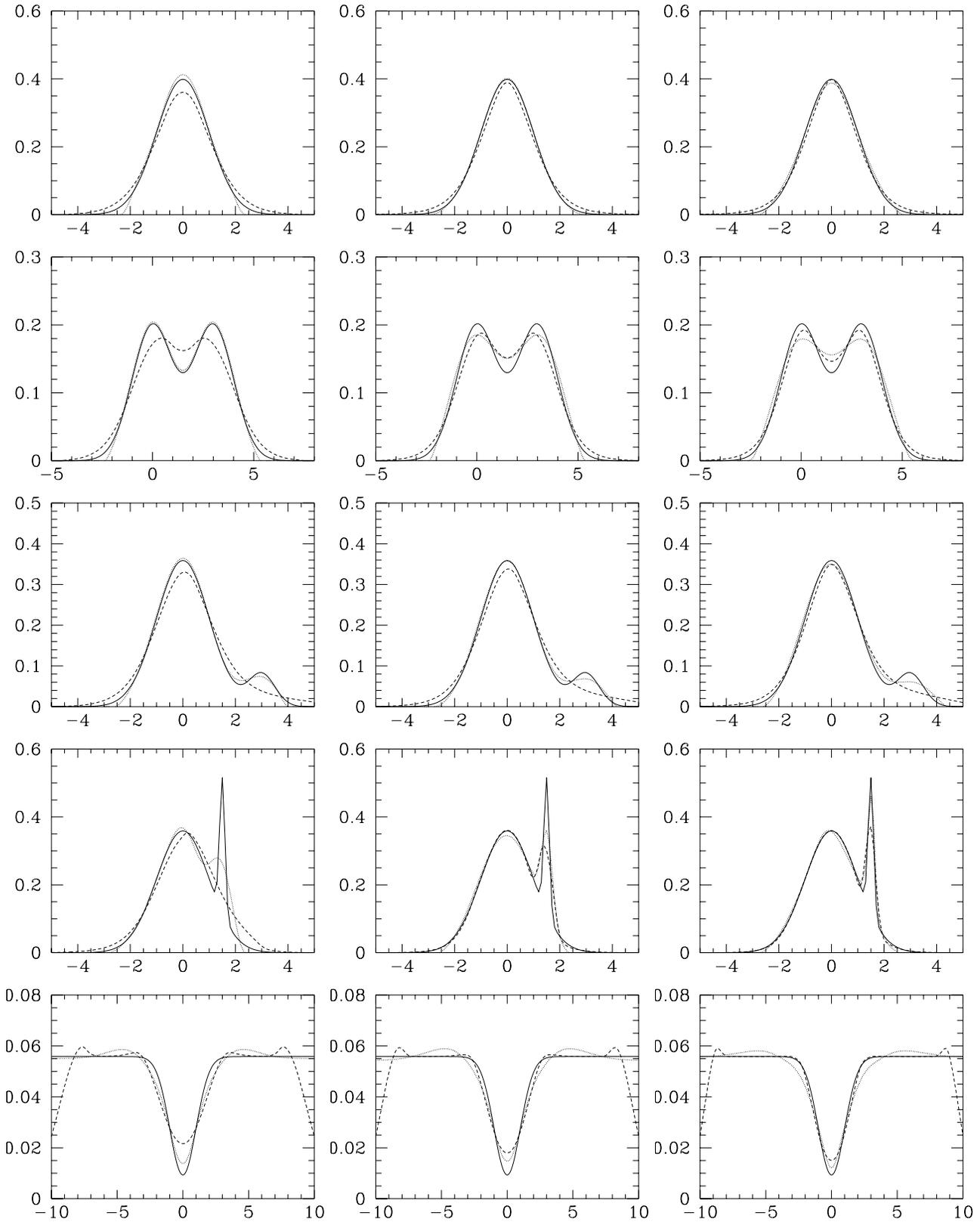,height=22.0cm}}
\caption []{Kernel and wavelet estimates on ``noiseless'' samples.
The solid line is the theoretical distribution, the dashed line stands
for the kernel estimate and the dotted line corresponds to the wavelet
solution. Examples A to E (see the text) are displayed from top to
 bottom. The number of data points increases from left to right.}
\end{figure*}

\begin{figure*}
%\picplace {22cm}
\centerline{\psfig{file=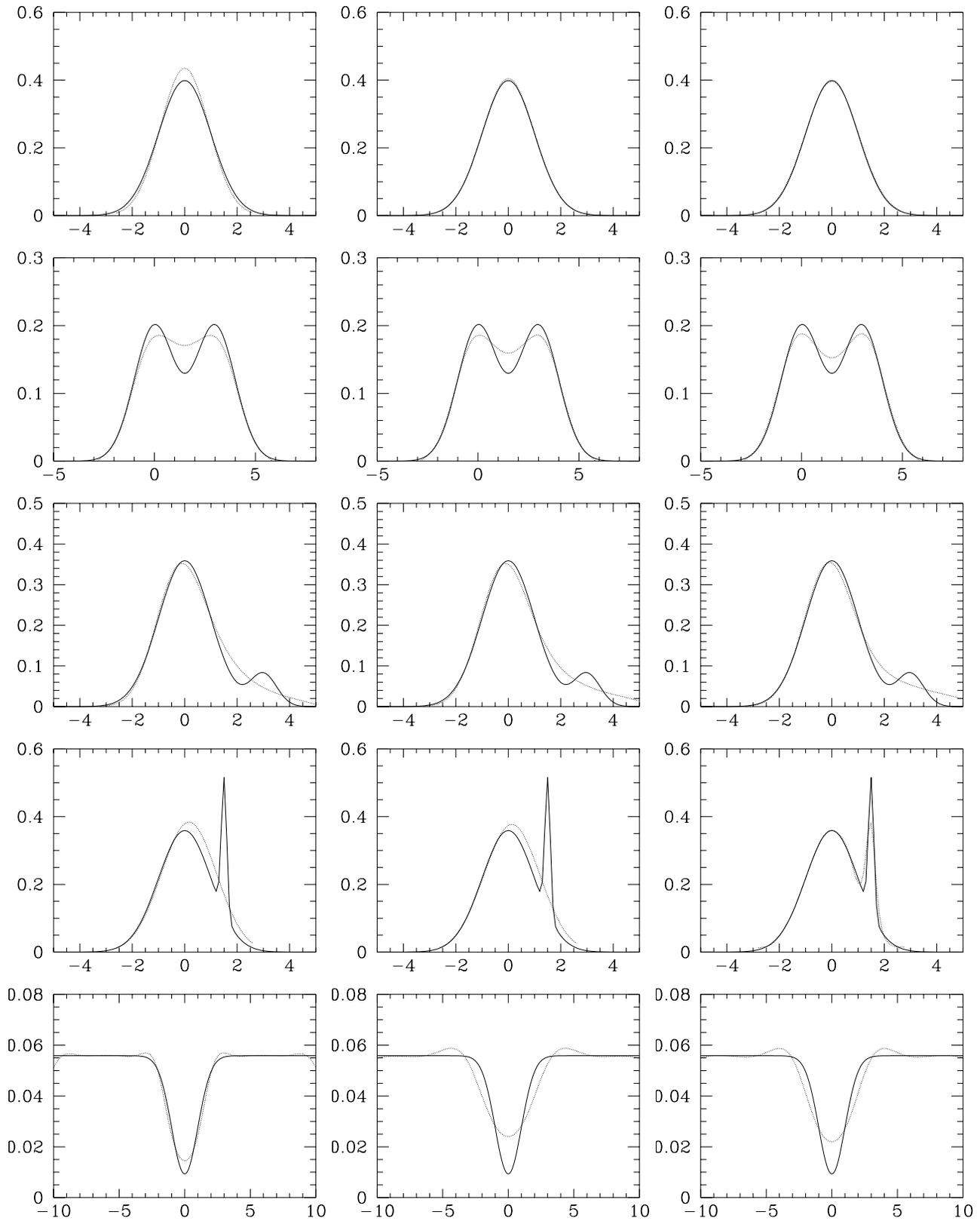,height=22.0cm}}
\caption []{MPL estimates on ``noiseless'' samples. The dotted line
corresponds to the MPL estimate. The graphs are sorted in the same way
as in Figure 1.}
\end{figure*}

\begin{figure*}
%\picplace {9.5cm}
\centerline{\psfig{file=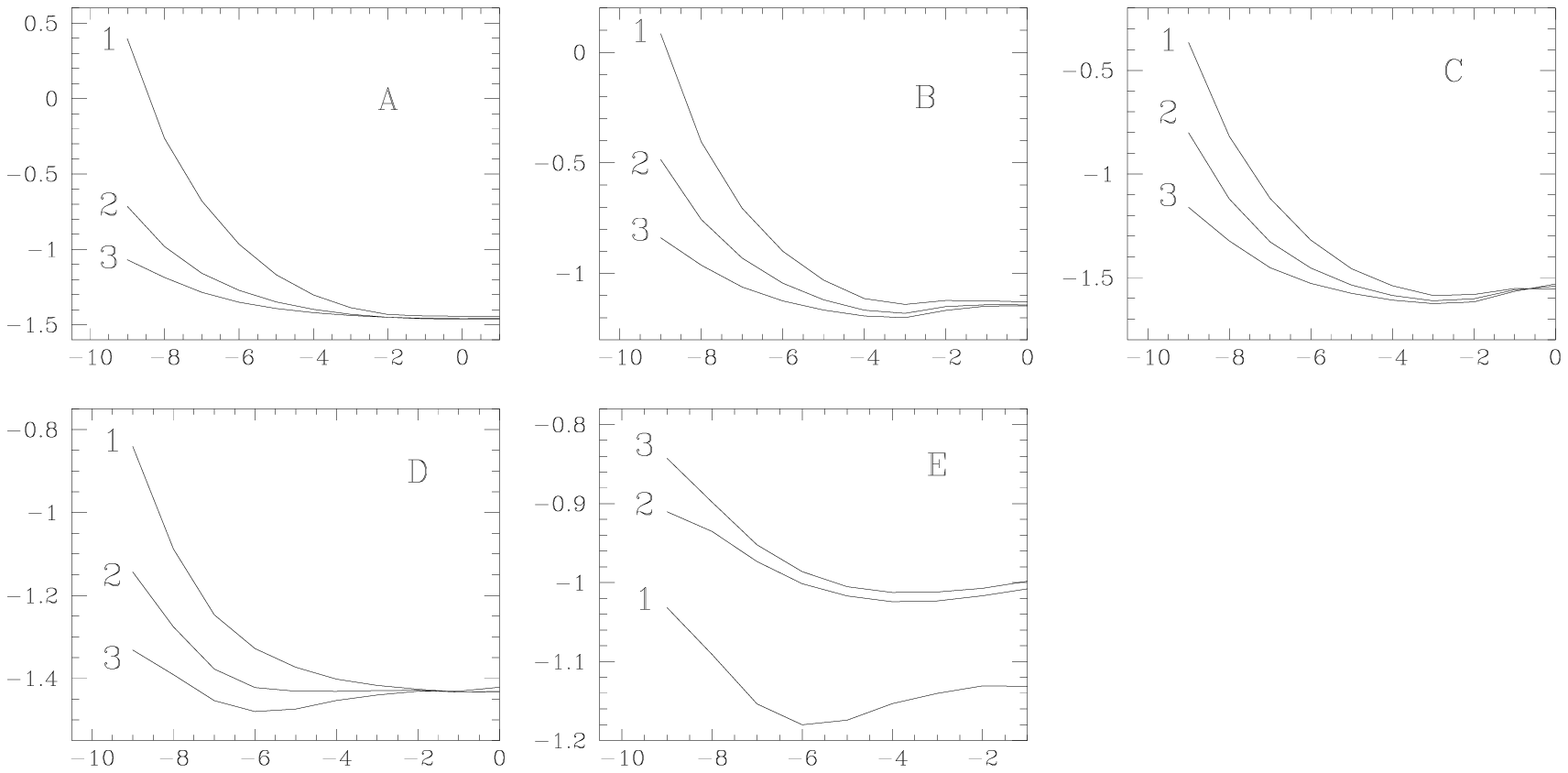,height=9.5cm}}
\caption []{The UCV functions related to the MPL estimator for 
the ``noiseless'' samples. Each graph is labeled according to the
corresponding example. The curve labels indicate the increasing sample
sizes.}
\end{figure*}

\subsection{Noiseless samples.}

First we considered the ``noiseless''  samples, which are generated by
transforming  a regularly    sampled uniform   distribution   into the
distributions of  the examples described  above.  The absence of noise
allows us to highlight the performance of the methods.  In Figure~1 we
show  the  estimations by  means    of  the kernel and     wavelet
estimators. In Figure~2 we report the MPL density estimates, while the
corresponding UCV curves are displayed in Figure~3.

The  comparison of the   whole set of   results  shows that  the three
methods detect and  retrieve most of the features  in more or less the
same way, especially  in the case of a great  number of  data.  The kernel
method yields quite accurate restored density functions in most cases,
with   the  noticeable  exception  of   example    C,  where the  small
superimposed  Gaussian is never  really detected.  The same difficulty
arises   for the MPL estimates.   On  the contrary, small features are
better  detected by  the wavelet method  than  by the  others. For
instance, only the wavelet method is  able to detect the small feature
of example C and the secondary peak of  example D when 100 data
points are involved.  The results of the MPL method are similar to those of the
kernel method.  Nevertheless,  it  appears that the  restoration  coming
from the  MPL method is  more accurate for  the  Gaussian tails of the
distributions, whereas  it fails to detect the   secondary peak of 
example D when  the sample size  is lower than 400.   \\ As for the MPL
estimates, it becomes  clear  by looking at  Figure~3  that there are  some
cases where it is not possible to find a minimum of the UCV; in fact,
only  monotonic decreasing curves  are sometimes obtained.  This means
that a  large value of the  penalization  parameter give  a good fit,
i.e. the MPL estimate becomes a Gaussian fit of the data (see \S~2.3).
Moreover, as discussed in the previous section. the MPL method suffers
from   its    dependency on  the  efficiency  of    the  algorithm of
minimization as well as from a computational time  which is much higher than
for the other two  methods.  These disadvantages prevent efficient
use of  the  method, especially  when   high  resolution is required.
Since the  overall performances of the  MPL  method appear to  be very
similar  to the other methods, we  decided to  investigate further only
the behaviors of the kernel and wavelet  approaches.  The MPL will be
referred to again only when analyzing some real data sets.  \\

Let  us now take a close look at the  general  behavior of both
methods     by means of     numerical simulations.  The trends and  subtle
differences between the kernel  and wavelet results will  be explained
by reference to their underlying mathematical definitions.

\begin{figure*}
%\picplace {22cm}
\centerline{\psfig{file=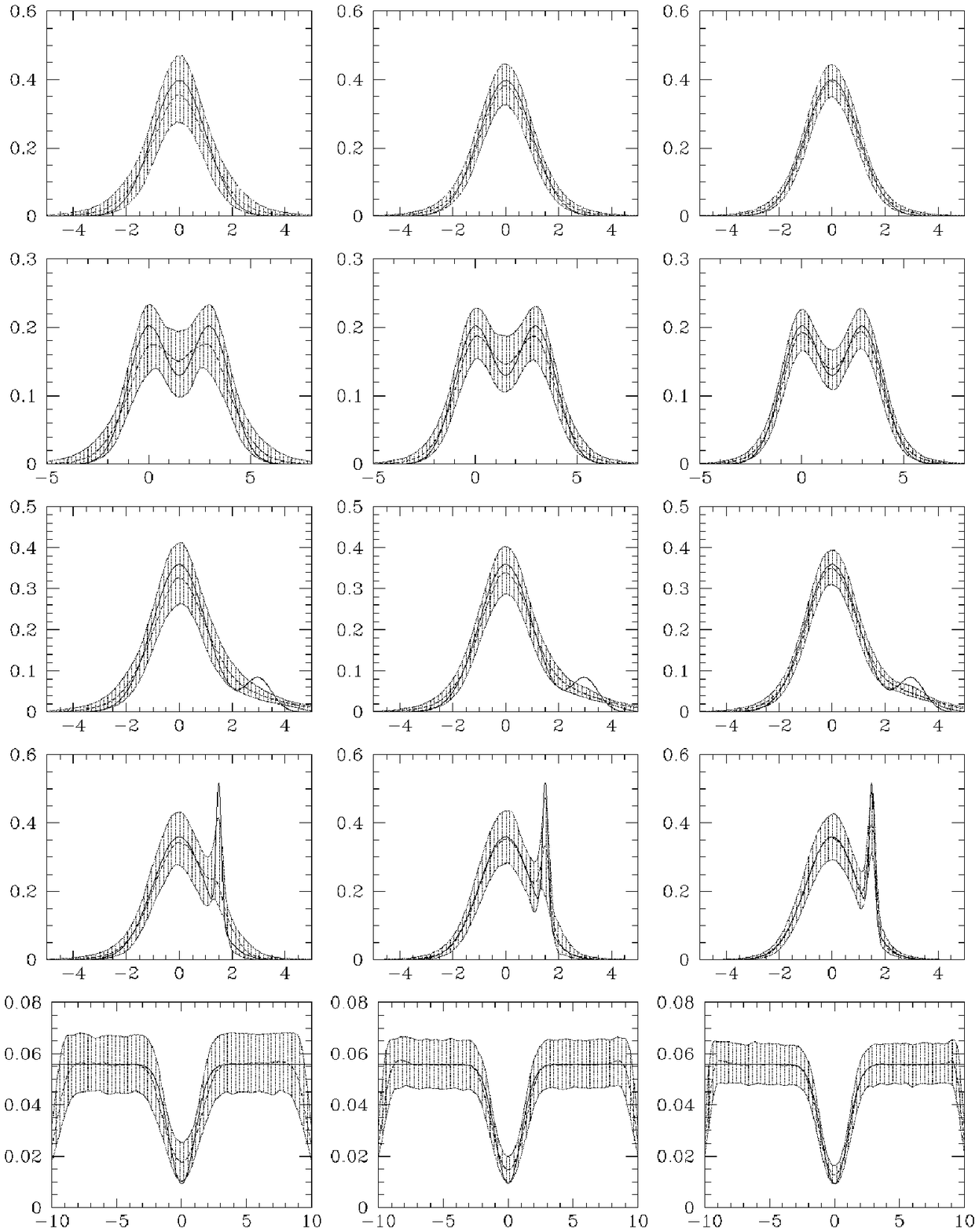,height=22.0cm}}
\caption []{Kernel results from numerical simulations. The graphs
are sorted in the same way as in Figure 1. Solid lines represent
the theoretical distributions; the hatched area is limited by the 10 and
90 percentiles of the results while the dashed line stands for the
median solution.}
\end{figure*}

\begin{figure*}
%\picplace {22cm}
\centerline{\psfig{file=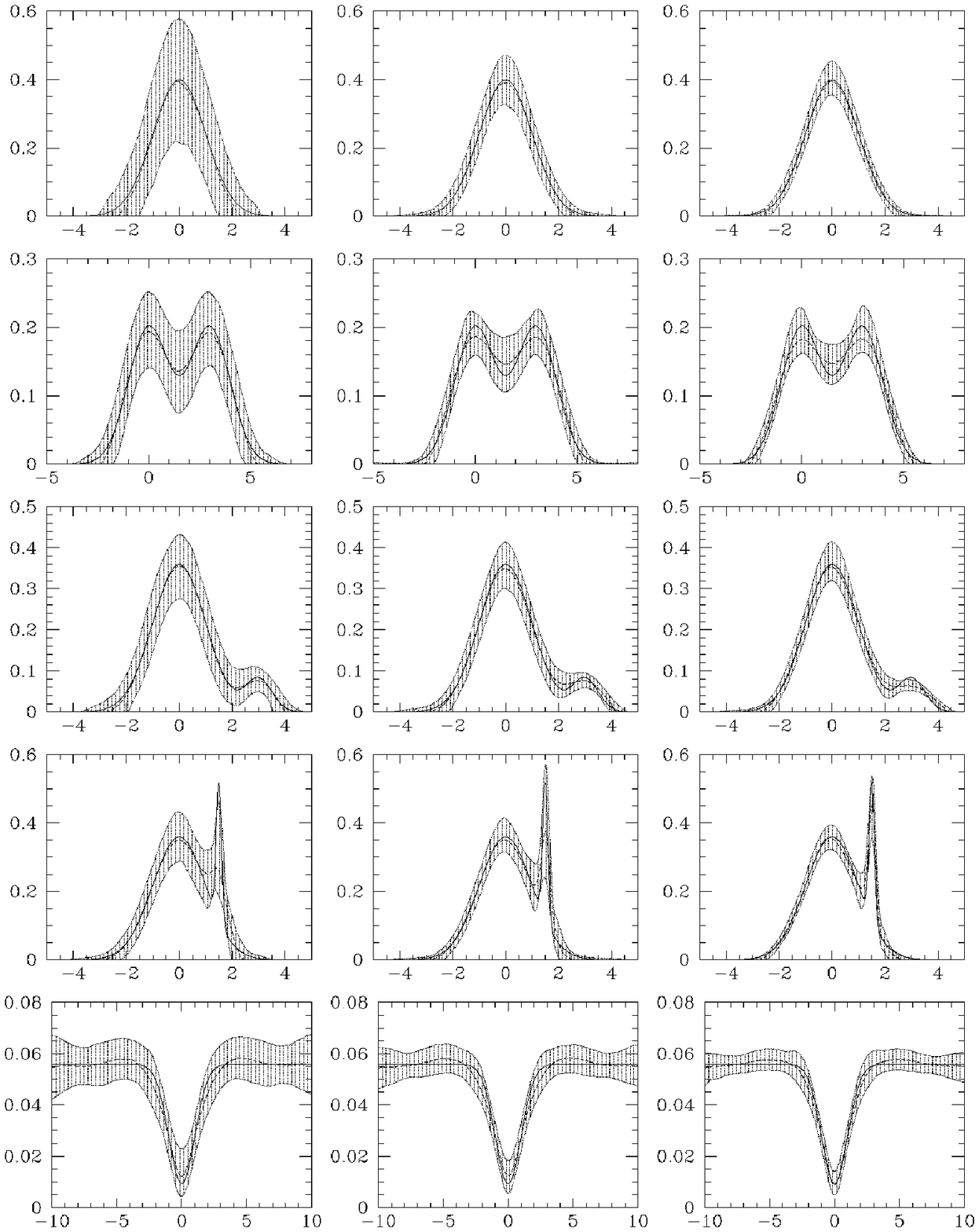,height=22.0cm}}
\caption []{Wavelet results from numerical simulations. Definitions
are the same as in Figure~4.}
\end{figure*}

\subsection{Statistics.}

We performed 1~000 simulations for each case in order to estimate
the variance of  the estimated density functions,  which  is linked to
the intrinsic variations in the experimental data set.

In  order to compare the  two density  estimations,  we chose to
evaluate the results    on a grid    of 50  points.  The   theoretical
functions  (solid  line), the median curves  of  the estimates (dashed
line) with their 10 and 90 percentiles (hatched band), which represent
a measure of the variance of the estimate, are displayed for each case
in    Figures~4  and~5   for    the  kernel  and  wavelet  estimators,
respectively.

These curves show  local  agreement between  the  estimates and the
true distributions.  We decided    to get    quantitative information
about the global quality of the solutions by evaluating the integrated
square error for the estimate of each simulation according to the formula:
\begin{equation} 
{\rm ISE} = \frac{1}{50}\sum_{i=1}^{50} (\hf_i-f_i)^2.  
\end{equation} 
The distributions of this quantity for the two estimators are displayed
in Figure~6.  We report the  ISE values for the ``noiseless'' estimate
in Table~1.

One of the aims of density estimations from discrete data is structure
characterization. Information, viewed in  terms of basic structure  parameters
with  respect to the  true values,  is  provided in Table~2.  It gives
positions and amplitudes for the peaks which are present in the median
estimates. The errors relate to the grid step for the positions 
and to the 10-- and 90--percentile values for the amplitude.

\subsection{Comments.}

First  of  all, our study shows   that both kernel-  and wavelet-based
density estimators recover different parent distributions quite  well,
though some  differences in  efficiency  can be noticed; moreover,  in
most cases the accuracy of the kernel and wavelet estimations increases
with the  number  of points,  while the  variance  decreases.   Let  us
examine in  detail the different   examples in order to describe  the
fine  behavior of these estimators,  which require approximately the same
computational resources.\\

{\em Example A.}\\

 When  dealing  with the  experimental set involving   a low number of
points, we first notice that  the variance is  larger for the  wavelet
estimate than for the kernel estimate. In fact, the wavelet transform is as
sensitive to  voids as to clustering in   the data distribution.  In
the case  of few data, significant voids  generated by random fluctuations
in the numerical set  are  frequently detected. Therefore, the  analysis
often  ends up  with several  small  clumps instead of  a single large
cluster with a Gaussian distribution. This increases the final variance of the
result.   However, the  median  curve and  the  ``noiseless'' estimate
agree fairly well with the  parent distribution, except for the tails.
In  fact, we  decided to consider  wavelet  coefficients computed with
fewer than  three points as meaningless  for statistical reasons. Since
there is a  low probability of  having  experimental data points in  the
tails, this explains why these features are missing both in the median
and  in the ``noiseless''  estimates. Cutting  the  tails is a general
behavior of our wavelet-based density estimates.

On the  contrary, the kernel  solution  presents wider tails  than the
parent distribution.  Wide kernel functions are  in fact associated with
every point in low density regions (see eq.~\ref{eq_kfun}). Thus, as a
consequence of  normalization, the kernel estimate departs from the
true function in the central region in the case of restricted sets of data
points.  These trends  are verified  for  every example in  our study.
Further  information   is provided by   Figure~6  which shows that the
global   agreement  with the  theoretical  function  is better for the
kernel than for the  wavelet estimate when  noisy data are considered.
Voids due to large statistical fluctuations are indeed not detected by
the  kernel estimator. This  characteristic of  the  kernel  method is
obviously relevant when regular distributions are sought for, but it
introduces a bias if the  genuine distribution  exhibits such holes  as
shown in the following examples.

Whit an increase in the number of points, both methods appear to give
quite similar results. The ISE distributions still differ, owing to
the differences in sensitivity of the estimators to voids.  This
indicator also shows in a prominent way the ability of the kernel to
reproduce almost perfectly the parent Gaussian distribution no matter
what the experimental set is.  But this disparity mostly disappears
when the ``noiseless'' set is considered; thus the wavelet estimator
has a clear advantage, especially at a low mean number (see Figures~5
and~6).\\

{\em Example B.}\\

If we analyze two identical  close  Gaussians,  it appears  that the
behavior of the two estimators is quite similar,  both from the point
of view of  local variance and of the ISE distributions.  This is a
general result which is true also for  the following examples.  However, in both
ideal ``noiseless'' and experimental situations, the results show that
the wavelet estimator is more efficient  in the case of few events,
and that this  superiority  vanishes  when  the number of
points increases.

The explanation is easy.  In the case of large data sets, the contrast
between high and low density regions is reduced, and fewer and fewer
simulations exhibit a strong gap between the two Gaussian peaks.
Therefore, the wavelet transform finds it more and more difficult to
exhibit the valley in the data distribution, and the median value and
the ``noiseless'' result accordingly increase between the two peaks,
since a crucial piece of information for the wavelet-based restoration
is missing.  Conversely, the efficiency of the kernel estimator in
detecting Gaussian peaks improves as the size of the data set grows,
leading to a better peak-to-peak contrast. \\

\begin{figure*}
%\picplace {22cm}
\centerline{\psfig{file=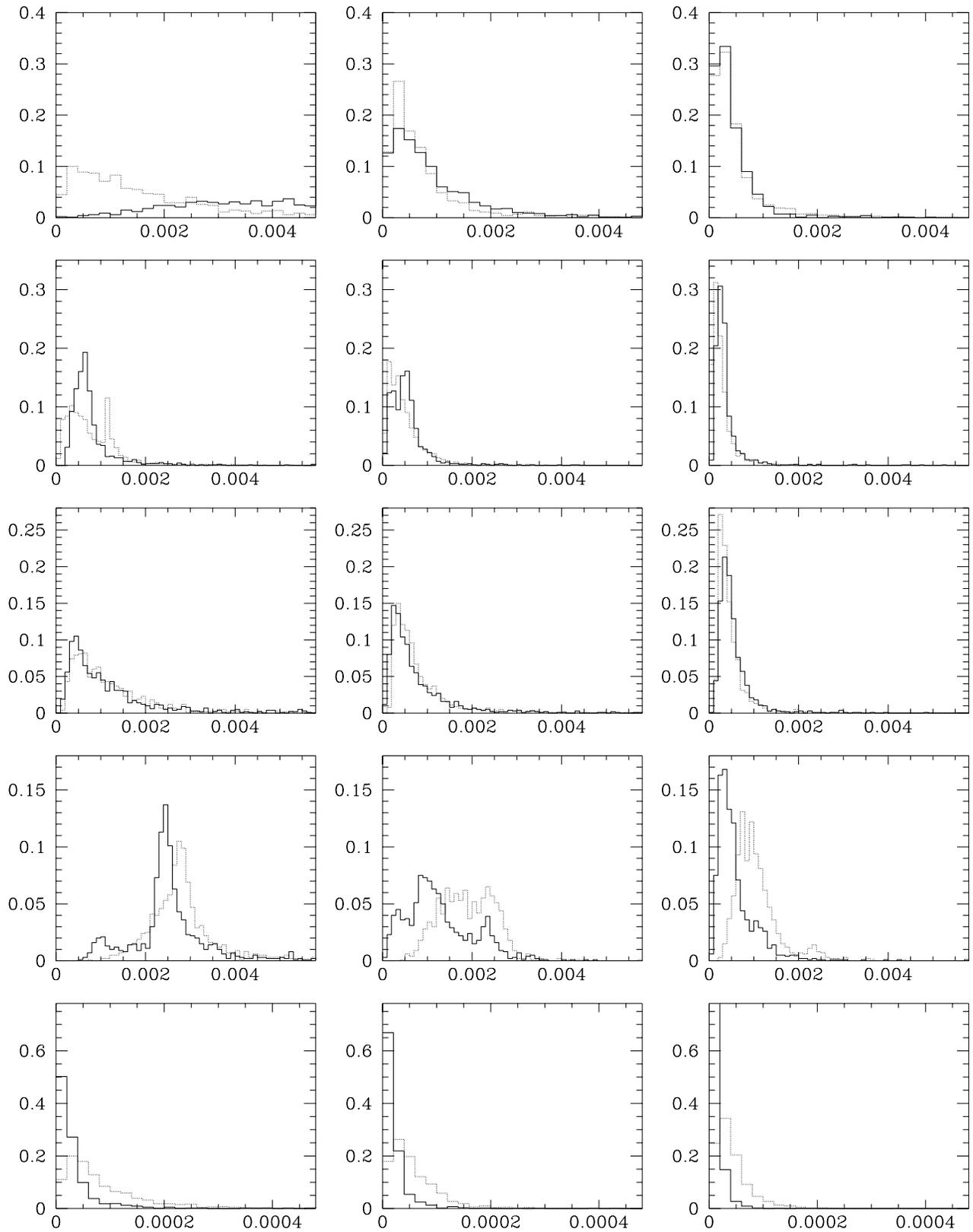,height=22.0cm}}
\caption []{ISE distributions for the kernel and wavelet estimates
corresponding to the samples considered. The graphs are sorted as
explained in Figure~1. The dotted  histogram corresponds to ISE values
for the kernel solutions, while the solid one displays the distribution
of ISE values for the wavelet estimates.}
\end{figure*}

%\documentstyle{article}
%\pagestyle{empty} 
%\begin{document}
\setcounter{table}{0}
\begin{table}
\caption[]{ISE values for kernel and wavelet estimates on
the ``noiseless'' samples.}
%\scriptsize
\begin{center}
\begin{tabular}{crrr}
\noalign{\smallskip}
\hline\noalign{\smallskip}
\multicolumn{1}{c}{Ex.}&
\multicolumn{1}{c}{N}&
\multicolumn{1}{c}{Kernel}&
\multicolumn{1}{c}{Wavelet}
\\
\noalign{\smallskip}
\hline\noalign{\smallskip}
A  & 30 &3.05$\cdot 10^{-4}$  &  9.14$\cdot 10^{-5}$  \\
   &100 &1.48$\cdot 10^{-4}$  &  7.20$\cdot 10^{-6}$  \\
   &200 &1.05$\cdot 10^{-4}$  &  4.53$\cdot 10^{-5}$  \\
B  & 50 &2.58$\cdot 10^{-4}$  &  1.86$\cdot 10^{-5}$  \\
   &100 &1.18$\cdot 10^{-4}$  &  8.43$\cdot 10^{-5}$  \\
   &200 &7.26$\cdot 10^{-5}$  &  1.26$\cdot 10^{-4}$  \\
C  & 50 &3.92$\cdot 10^{-4}$  &  3.21$\cdot 10^{-5}$  \\
   &100 &3.23$\cdot 10^{-4}$  &  3.03$\cdot 10^{-5}$  \\
   &200 &2.51$\cdot 10^{-4}$  &  7.30$\cdot 10^{-5}$  \\
D  &100 &2.58$\cdot 10^{-3}$  &  1.55$\cdot 10^{-3}$  \\
   &200 &1.11$\cdot 10^{-3}$  &  8.16$\cdot 10^{-4}$  \\
   &400 &5.28$\cdot 10^{-4}$  &  1.57$\cdot 10^{-4}$  \\
E  &-50 &1.46$\cdot 10^{-4}$  &  5.37$\cdot 10^{-6}$  \\
   &-100&1.14$\cdot 10^{-4}$  &  1.67$\cdot 10^{-7}$  \\
   &-200&8.90$\cdot 10^{-5}$  &  7.17$\cdot 10^{-7}$  \\
\noalign{\smallskip}
\hline
\end{tabular}
\end{center}
\end{table}
%\end{document}

{\em Example C.}\\

This example clearly exhibits some consequences of the general behaviors
pointed out just above.  The key result of  the test is that the small
feature on the  right side of the main  curve is recovered only by the
wavelet estimator.  The feature is even more evident when   a few number  of
points is  involved in the estimate,  the efficiency becoming lower as
the sample  size    increases as  pointed out  before.    Meanwhile,   the
asymmetry in the kernel estimate could be used to
deduce the presence of a feature otherwise missed.

This discrepancy  can be easily understood. It is very difficult for
the kernel estimator to detect small features as it relies solely on
the  related small clusters of points  to recover the  signal.  On the
contrary, the  wavelet estimator also detects  the presence of voids,
and such    information  is  of great  importance  when  broad small
structures are  sought for, which is the  present situation.  So it
appears that the wavelet estimator does not recover the secondary peak
only by relying on its points, but   rather  by  also detecting  the underdensity
which separates it  from the main structure.   The contrast diminishes 
as the density increases; this explains
why the   secondary peak is blurred  in   the last  high--density case
(cf. example B). \\

{\em Example D.}\\

A peaked small cluster has now to be  recovered within a main Gaussian
distribution.  The smoothing caused by the use  of kernel functions, as
well  as the ability  of the wavelet-based  method  to make use of the
gaps in the  data sets  are  also exhibited here.  In  fact, although
both estimators give correct  and similar results  when the number  of
data is   high enough to  define   both structures properly, their
respective behaviors are again different for a limited set of points.
The wavelet estimator succeeds in exhibiting  the secondary peak, even
if  its shape   parameters are  poorly  determined,  while  the kernel
estimate shows only a marked asymmetry for the ``noiseless'' sample or
a small  deviation  from the  pure Gaussian for  the experimental data
set.   The resulting variance is then  lower  for the wavelet estimate
than for the kernel one.

These facts are not surprising.  Both methods  are sensitive to strong
clustering and detect the secondary peak with increasing efficiency
as the size of  the sample increases.   But, as said before, the use
of kernel functions tends to smooth the data, so that small clumps are
erased and real small  voids are missed.  On  the other side,  the wavelet
transform   enhances and  makes use of   both  features, whatever their
scales may be.    This difference is  striking when  the sample  with the
smallest number of data is analyzed. \\

{\em Example E.}\\

We have now  to  deal  with a deep    hole located within a   constant
high-density   region.  As  shown   by  the  variances   and the   ISE
distributions, the wavelet estimate is  better for recovering the hole,
no matter what  the size of  the sample is. However,  the kernel method 
also does a good job when the sample is not too small.\\

%\documentstyle{article}
%\pagestyle{empty}
%\begin{document}
\setcounter{table}{1}
\begin{table*}
\begin{center}
\caption[]{Structure parameters.}
%\scriptsize
\begin{tabular}{crrrrcrrr}
\noalign{\smallskip}
\hline\noalign{\smallskip}
\multicolumn{1}{c}{Ex.}&
\multicolumn{1}{c}{N}&
\multicolumn{3}{c}{Location}&&
\multicolumn{3}{c}{Amplitude}\\
\cline{3-5}\cline{7-9}
\multicolumn{1}{c}{}&
\multicolumn{1}{c}{}&
\multicolumn{1}{c}{True}&
\multicolumn{1}{c}{Kernel}&
\multicolumn{1}{c}{Wavelet}&&
\multicolumn{1}{c}{True}&
\multicolumn{1}{c}{Kernel}&
\multicolumn{1}{c}{Wavelet}\\
\noalign{\smallskip}
\hline\noalign{\smallskip}
 A &      30 &   0.00  &     $-$0.10$\pm$0.10 & $-$0.10$\pm$0.10      &&
                 0.40  & 0.36$^{+0.11}_{-0.08}$&  0.39$^{+0.18}_{-0.17}$\\
   &     100 &   0.00  &    $-$0.10$\pm$0.10  & $-$0.10$\pm$0.10      &&
                 0.40  & 0.38$^{+0.06}_{-0.05}$ &  0.39$^{+0.08}_{-0.06}$ \\
   &     200 &   0.00  &     $-$0.10$\pm$0.10  & $-$0.10$\pm$0.10      &&    
                 0.40  & 0.39$^{+0.05}_{-0.04}$ & 0.39$^{+0.06}_{-0.04}$ \\
 B &      50 &   0.00  & 0.31$\pm$0.13  &  0.04$\pm$0.13  &&
                 0.20  & 0.17$^{+0.05}_{-0.03}$&   0.19$^{+0.06}_{-0.05}$    \\
   &         &   3.00  & 2.69$\pm$0.13       & 2.96$\pm$0.13&&
                 0.20  &  0.17$^{+0.05}_{-0.03}$& 0.19$^{+0.06}_{-0.05}$    \\
   &     100 &   0.00  & 0.04$\pm$0.13  &  0.04$\pm$0.13  &&   
                 0.20  &  0.19$^{+0.04}_{-0.03}$&   0.19$^{+0.03}_{-0.03}$ \\
   &         &   3.00  &  2.96$\pm$0.13       & 2.96$\pm$0.13&&   
                 0.20  &   0.19$^{+0.04}_{-0.04}$&   0.19$^{+0.04}_{-0.03}$  \\
   &     200 &   0.00  &  0.04$\pm$0.13  &  0.04$\pm$0.13  &&
                 0.20  &   0.19$^{+0.03}_{-0.03}$&   0.18$^{+0.05}_{-0.02}$ \\
   &         &   3.00  &  2.96$\pm$0.13       & 2.96$\pm$0.13&& 
                 0.20  &   0.19$^{+0.03}_{-0.03}$&   0.18$^{+0.05}_{-0.02}$  \\
 C &      50 &   0.00  &       $-$0.10$\pm$0.10 & $-$0.10$\pm$0.10&&
                 0.36  &  0.32$^{+0.08}_{-0.07}$&  0.35$^{+0.08}_{-0.08}$ \\
   &         &   3.00  &  ~$\cdots$&   2.96$\pm$0.10      &&
                 0.08  &  ~$\cdots$&   0.08$^{+0.03}_{-0.03}$   \\
   &     100 &   0.00  &        $-$0.10$\pm$0.10 & $-$0.10$\pm$0.10&&    
                 0.36  &     0.34$^{+0.06}_{-0.05}$&  0.35$^{+0.07}_{-0.05}$ \\
   &         &   3.00  &  ~$\cdots$&   2.96$\pm$0.10&&
                 0.08  &  ~$\cdots$&   0.07$^{+0.02}_{-0.02}$       \\
   &     200 &   0.00  &         $-$0.10$\pm$0.10 & $-$0.10$\pm$0.10&&    
                 0.36  &     0.35$^{+0.04}_{-0.04}$&  0.35$^{+0.06}_{-0.03}$\\
   &         &   3.00  &  ~$\cdots$&   2.96$\pm$0.10&&
                 0.08  &    ~$\cdots$&   0.07$^{+0.02}_{-0.01}$          \\
 D &     100 &   0.00  &   $-$0.10$\pm$0.10 & $-$0.10$\pm$0.10&&   
                 0.36  &   0.34$^{+0.09}_{-0.06}$&  0.35$^{+0.08}_{-0.06}$  \\
   &         &   1.50  &  ~$\cdots$&   1.53$\pm$0.10&&
                 0.52  &  ~$\cdots$&   0.26$^{+0.10}_{-0.06}$     \\
   &     200 &   0.00  &   0.10$\pm$0.10 & $-$0.10$\pm$0.10&&  
                 0.36  &   0.35$^{+0.08}_{-0.07}$&  0.35$^{+0.06}_{-0.03}$  \\
   &         &   1.50  &  1.53$\pm$0.10 &   1.53$\pm$0.10&&
                 0.52  &  0.33$^{+0.14}_{-0.15}$&   0.37$^{+0.20}_{-0.14}$  \\
   &     400 &   0.00  &   $-$0.10$\pm$0.10 & $-$0.10$\pm$0.10&&
                 0.36  &   0.36$^{+0.06}_{-0.07}$&  0.36$^{+0.04}_{-0.04}$  \\
   &         &   1.50  &  1.53$\pm$0.10 &   1.53$\pm$0.10&&
                 0.52  &  0.39$^{+0.09}_{-0.09}$&   0.45$^{+0.09}_{-0.10}$  \\
\noalign{\smallskip}
\hline
\end{tabular}
\end{center}
\end{table*}
%\end{document}

One can notice that the tails of the Gaussian hole are somewhat larger
in the wavelet-based estimate than in the kernel one, and that the two
small bumps which delineate the boundaries of  the void are higher for
the wavelet solution.  These effects  are related to rapid  variations
in the shape  of the  distribution  and are very   evident in the case  of
discontinuities.  Both effects are  due to the  shape of the analyzing
wavelet function which must be designed to yield zero-valued coefficients for a
uniform distribution (see Figure~9 in  the Appendix).  In such a case,
wavelet  coefficients  are    indeed equal   to   zero,  since positive
contributions equal negative ones.  But, as locations closer to a hole
are  examined,  the density of points  decreases   in one  part of the
negative area of the  function,  yielding some positive values  before
ending with  the negative ones denoting the  void.  Such artifacts are
intrinsic  to    the wavelet method   when  a   background is   to be
considered.  This concerns obviously voids but also peaks superimposed
on  a constant    background~:  two symmetrical   positive  or  negative
contributions appear, respectively.    However, this effect  is strong
enough to generate significant structures and is a problem for further
analyses only when the main  structure is strongly contrasted with respect
to the background or when the signal itself  is very irregular.  While
negative features are unrealistic and  can be easily thresholded by using
a positivity  constraint  (see  eq.~\ref{eq_rest}), only  a  dedicated
processing of  the  wavelet-based density  estimate can allow  one  to
remove them in a systematic way.  Guidelines for doing so are given in the
next  section.   Nevertheless, most  of   the  cases  of  astronomical
interest concern  peaks   located inside a  low  and   nearly constant
background (cf. introduction), so  that the quite simple wavelet-based
method  described  here  can  be  used  with   great  advantage in  most
situations without any particular difficulty.\\

\begin{figure*}
%\picplace {12cm}
\centerline{\psfig{file=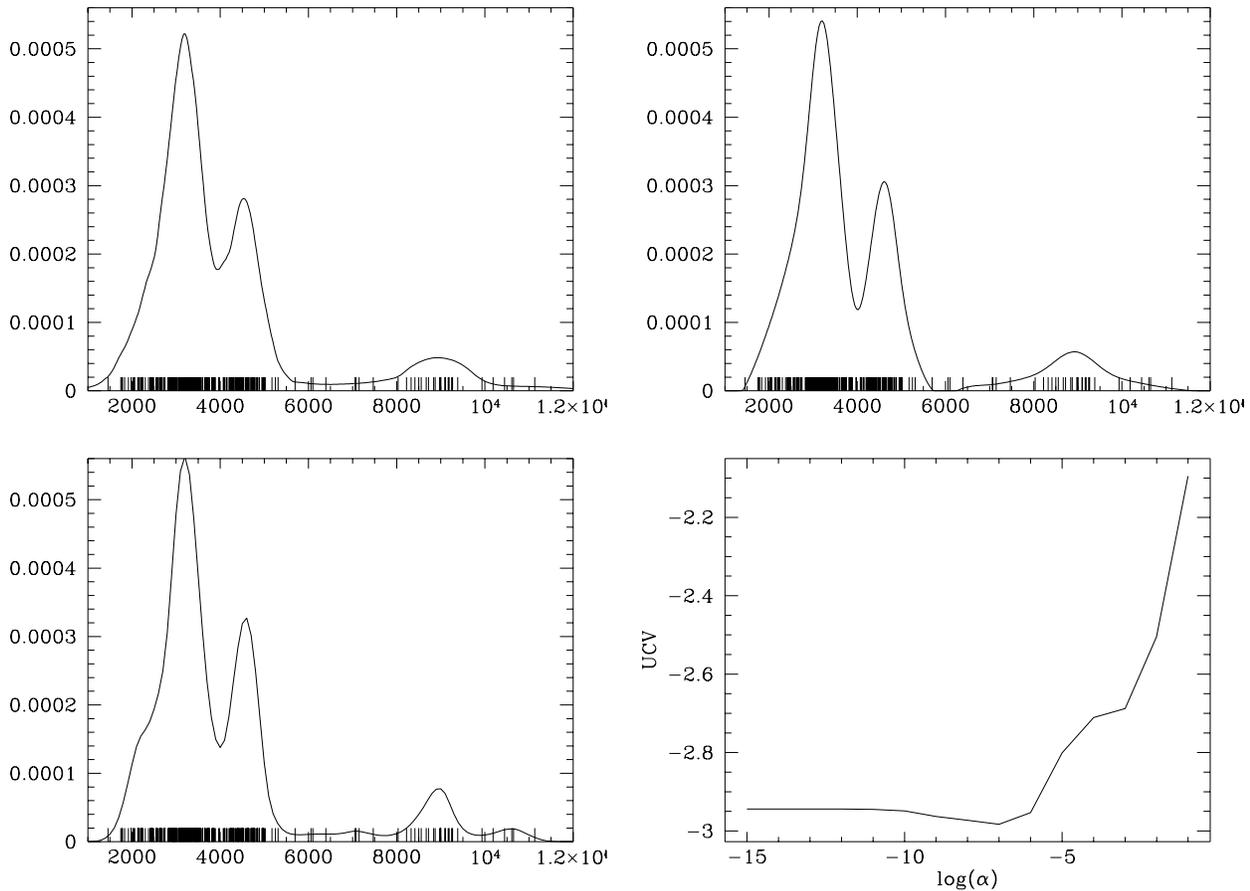,height=12.0cm}}
\caption []{Analysis of the redshift distribution of the A3526 galaxy 
cluster.  At top are displayed the kernel (left) and wavelet (right) 
estimates. At bottom is given the MPL solution with  the UCV function 
of the estimator. At the base of each estimate, the bars stand for the 
observational data. The unit of the x-axis is km s$^{-1}$.}
\end{figure*}

\subsection{General remarks.}

These examples enable us to make some general remarks about the way the
kernel and wavelet density estimators analyze  a discrete catalogue in
order to recover the underlying density function. \\

Both estimators appear to give very similar results in most cases.  In
fact, the kernel makes use of a  smoothing function whose size depends
on  the local  density, while wavelets  select the  scale which is 
most appropriate  for defining  the local   signal.    However, kernel
estimates fail  to detect unambiguously  faint structures superimposed
on a larger component (example C) or poorly defined groups (example D,
case  1). Conversely,   wavelet-based solutions appear   to  
find it difficult to accurately disentangle merged structures of comparable
scale   when  the sample    size is  large  (case   3,  examples  B \&
C). Moreover, the sensitivity  of wavelets to voids generates negative
values of the density  which have to  be thresholded, thereby inducing
discontinuities   at  the   zero-crossing    locations.  These   voids
correspond to  strong gaps in  the data or  to regions with fewer than
the  minimum   number of points   required  to  compute  a meaningful
significance level.  Finally, in  all  the examples, wider tails   are
generated by kernel  estimates   than by wavelet  ones.  Wide  kernel
functions are  summed  together in  low  density   regions  where  no
significant wavelet coefficients are usually found.

The kernel estimator takes  into account  only  the presence  of data,
whereas  the wavelet  estimator  relies    on  local over--   and
underdensities detection to restore the  density function.  Therefore,
in the case of  a  restricted  set  of data  or  when  dealing with  very
different mixed distributions,  wavelets are more suitable than kernel
functions, since   two  kinds of information   about  the  local density
contrast  can  be used.   When   these  density  contrasts  are   less
prominent,   the wavelet  method may  be    less  efficient  than  the
kernel-based estimator.  For instance, this may  occur as gaps between
close  distributions disappear, owing  to  the increasing size of the data
sample.  On the contrary, the efficiency of the kernel solution always
increases with the number of data points.\\

With regard to  the void detection, the  wavelet estimator performs better than
the kernel  one.  But the  solution  obtained has two  small symmetric
artifacts which may  cause false detections and  have to be removed to
allow  fully automated analyses  (this is  also true for
the other two estimators).  An  iterative solution is available within
the wavelet framework, since   this   method enables one to     restore
separately each structure which constructs  the density distribution  function
(see  Ru\'e \& Bijaoui 1997, Pislar  \et 1997).  The solution
relies  on a structure-based   description  of the  signal.  The  main
component has  first to be  detected  and restored by using its  wavelet
coefficients.   The  obtained structure is  then   subtracted from the
zero-th order  density   estimate (see eq.~\ref{eq_zero}),  and  a new
search  for structures is  performed until no more significant wavelet
coefficients  are  detected.  Alternate  restorations  are  needed  to
accurately  determine the shape parameters  of close structures. In this
way, the density estimate  may be computed as  a sum of genuine single
structures.

In  a  forthcoming  paper we plan  to   apply  this  procedure   to
two-dimensional sets of  data to get  a better analysis of  the galaxy
distribution within  galaxy  clusters.  In  fact, apart from a  continuous
density   estimation,  we  are  mostly  interested    in  an  accurate
description  of our  data   sample in  terms   of structures~: cluster
identification, evidence   for  subclustering,  shape parameters  with
respect to theoretical models, etc.  Nevertheless, Table~2 shows that
the available information is already  good enough to recover the  main
parameters of the   underlying theoretical Gaussians involved  in  our
examples, both for wavelet and for kernel estimators.

The  kernel-based   method could   also  be    improved with a   better
identification  of the optimal smoothing parameter  by means of a more
efficient data-based  algorithm. This will result in a   better  density
estimate from  the  point of view  of either  the resolution  or the
significance of the solution.

The same remark also holds for the MPL technique. However, the use of
a more efficient minimization algorithm would be  also needed in order
to make  this method faster and to  improve its resolution.  This is a
necessary step for applying the method to multivariate cases.

\begin{figure*}
%\picplace {12cm}
\centerline{\psfig{file=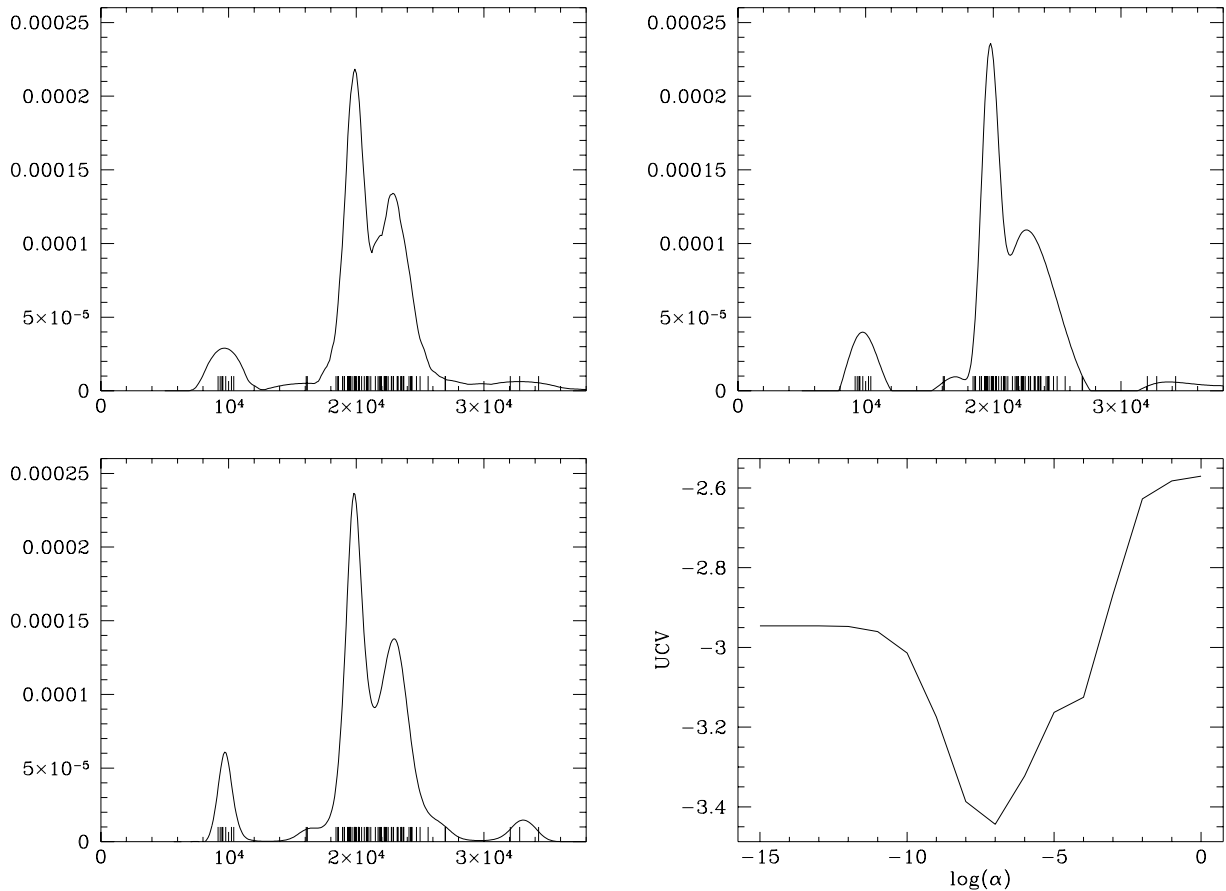,height=12.0cm}}
\caption []{Analysis of the redshift distribution of the {\it Corona
Borealis} sample. Definitions are the same as in Figure~7.}
\end{figure*}

\section{Tests on astronomical data sets.}

In  the  present section we   apply the   three methods   to two  real
one-dimensional astronomical data samples~: 301 measured redshifts for
the cluster  of galaxies Abell~3526 (Dickens  \et 1986) and a
redshift sample of 82 galaxies in  the region of {\it Corona Borealis}
(R\"oder 1990).\\ The  Abell~3526 cluster  was already considered   by
Pisani (1993) in order to study the performance of the adaptive kernel
method.  Abell~3526  is a bimodal  cluster  in the redshift space (see
e.g.  Lucey \et 1986) and it provides  us with an example of
moderate complexity, intermediate  between examples B and  D. Figure~7
shows the kernel  and wavelet estimates,  as well as the MPL  estimate
together with  the UCV  function  allowing one to  obtain the  optimal
penalization  parameter. The bars at the  base of  the plots stand for
the observed redshifts. The second sample  is studied in order to make
a  comparison with  the results  of  Pinheiro \& Vidakovic  (1995) who
developed   another    wavelet density  estimator    based  on  a data
compression approach.  Our estimates with the kernel, wavelet, and MPL
methods are shown in  Figure~8, as well as  while the UCV function for
the MPL estimator.\\

As expected  from  the numerical simulations,  the three  methods give
consistent results in  both  cases. The differences  are similar to those
exhibited in the studies described in  \S~3.4.  In fact, the use
of  the wavelet  estimator  results    in sharper  and  more   compact
structures when compared to kernel estimates, and  it may allow one to
detect small features otherwise missed (e.g. the peak located at $\sim
1.8\ 10^4$~km/s  in   Fig.~9).  But,  as usually,  discontinuities  at
zero-crossing locations occur  in these wavelet-based estimates.   The
MPL and  kernel solutions are defined as positive, but only MPL
estimates can exhibit regions  of null density for  local voids in the
data.  Hence, the MPL  estimates differ from  the kernel  solutions by
yielding structures with a somewhat smaller support and regions of low
density,  similar to those restored  in the wavelet-based approach, but
without discontinuity problems (cf. Figure~1).

When dealing  with the A3526   data, three structures  are detected, in
agreement with previous studies.  The bimodality of the cluster is
confirmed, as  well as the existence of  a background group 4~000 km/s
away from the main structures. The three methods  we have used succeed
very   well in separating  the two   peaks defining  the  body of  the
cluster. The significance of both results is at least at the 3.5 sigma
level (cf. the  threshold  applied  to the wavelet  coefficients  with
respect to their statistical significance).

As for the {\it Corona Borealis} sample, our  results indicate that the
distribution of  redshifts  is composed of a   foreground group, a complex
central structure  and a background  population without any clear  sign of
clustering.  The central structure  is mainly bimodal, but the overlap
between  the two peaks  with different heights  is  greater than in the
A3526 case. Thus, no  firm conclusion about the  shape of  their profiles
can be reached until  alternate restorations have been performed  (see
\S~3.5). A small bump before the body of  the distribution denotes the
presence   of an  isolated  pair of  galaxies.  With   respect to  the
estimate  of Pinheiro \& Vidakovic (1995),  our solutions are smoother
but look similar, except for the smaller background peak of the central
structure. According  to   the  previous  density estimation,    this
secondary component is itself bimodal and much more clearly separated
from the  main  peak.   This  difference  comes  from   the underlying
strategies.  We are looking for a description  in terms of significant
structures,  whereas an efficient data  compression  is sought for in
the  other algorithm.  So   it  appears  that Pinheiro  \&  Vidakovic's
estimate follows  the data  more closely than  ours, which  is not the
optimal solution from the density estimation point of view.

\section{Conclusions.}

In this paper we  have studied the  efficiency of three recent density
estimators, namely   the adaptive kernel   method introduced by Pisani
(1993),  the  maximum penalized   likelihood described  by  Merritt \&
Tremblay (1994),  and our own  wavelet-based technique.  Wavelets have
already been used to recover  density estimations from a discrete data
set (Pinheiro \&  Vidakovic  1995), but with a  thresholding strategy
involving the  average energy of the wavelet  coefficients  at a given
scale.   Here the thresholding  is  defined with  respect to the local
information content, which enables us to obtain  a better estimate from
the statistical point of view.   Several dedicated examples were
used    to compare these   methods by  means    of extensive numerical
simulations.  These tests were chosen in  order to cover several cases
of   astronomical  interest  (cluster   identification,  subclustering
quantification, detection of voids, etc.).

Both   experimental  and ``noiseless''   simulations indicate that the
kernel  and the wavelet  methods can be  used with reliable results in
most cases.  Nevertheless, it appears that the best solution is always
provided  by the wavelet-based   estimate  when few  data  points  are
available.  The situation is more intricate when  the number of points
is  large.   Whereas the adaptive kernel   estimator  fails to clearly
detect  a small broad structure superimposed  on a larger  one, it can
yield  better results for   separating  two close, similar  structures.
As regards  void   detection, the wavelet    estimate gives  more confident
results,  but exhibits wider tails  and higher spurious  bumps on both
sides of the underdensity.

Accounting for the genuine voids properly   in the
experimental distribution   appears  to  be  the main  reason   for the
differences between the two approaches.   The kernel method associates
a  smoothing function to  each  data point  and the information coming
from gaps  in  the data  is  not  explicitly  used for recovering  the
density function. On the contrary, the wavelet  transform is  able to
detect  both overdensities and  underdensities in the  same way.  This
approach is therefore more efficient in analyzing data sets where both
highly contrasted features occur, which is especially the case in poor
samples.  When the  contrast is  reduced owing to  an  increase in  the
number of data, both methods give similar estimates.

The MPL method  performs  as well  as  the kernel-  and  wavelet-based
approaches,  as indicated by the  ``noiseless'' simulations. It appears
that  the results are somewhat intermediate  between those obtained by
means of the other two methods. However,  it strongly suffers from the
computational  cost  of   the minimization  algorithm  adopted,  which
prevents its use for large data sets.

The three methods were   applied to two redshift catalogues   of
galaxies which had  already been used to check  the efficiency of the
kernel method and  of   another wavelet approach,  respectively.   The
bimodality   of the A3526  galaxy  cluster   is displayed  by  all the
methods, as well  as the existence of a  background group of galaxies.
Both results confirm  the previous claims.   A redshift sample  from a
survey of     the   {\it Corona   Borealis}   region  was  also  
analyzed.  There  also, all  the  estimates are consistent, mainly indicating
a more intricate  bimodality  than in the  A3526 sample.   When
compared to    the alternative  wavelet-based  algorithm  proposed  by
Pinheiro \& Vidakovic (1995),  our solutions indicate that the wavelet
approach we have developed  performs better from  the point of view of
density estimation.\\

In conclusion, taking into account the computational inefficiencies of
the MPL method, both the kernel and wavelet  methods can be used to obtain
confident estimates of the underlying density related to discrete data
samples.  Wavelet  solutions  are to    be  preferred in  searching   for
subclustering,   especially  in  the case   of few   data   points.  Kernel
estimations  are more robust and perhaps  easier to implement.  Hence,
this approach appears to be very useful for arriving at reliable  solutions,
if it does not matter that some small--scale details may not be detected.
However,  only  the wavelet  approach  enables one to
naturally  decompose the restored density function  in terms of single
structures.  Such  decomposition is one  of the main goals to be achieved
for a deeper understanding of the dynamical status of galaxy clusters.

\begin{acknowledgements}
     We   are grateful    to Frederic   Ru\'e   for  many  stimulating
     discussions about  the  subtleties of  the wavelet   restoration
     algorithm.   F. D. wishes to   thank the {\it Observatoire de  la
     C\^ote d'Azur} for its kind hospitality and Prof. F. Mardirossian
     for his friendly support.
\end{acknowledgements}

\appendix

\section{Kernel estimator}

The  adaptive kernel estimators  imply  the use  of a  local smoothing
parameter $h_i=\lambda_i  h$ (see eq.~\ref{eq_lambda}). The quantities
$\lambda_i$ are   proportional to the  local density  at  location
$x_i$.  They are defined in Silvermann (1986) as~:
\begin{equation}
\lambda_i=\left[\frac{f_p(x_i)}{(\prod_j f_p(x_j))^{1/N}}\right]^{-\alpha},
\label{eq_kfun}
\end{equation}
where  $f_p(x)$  is a pilot estimate  of  the density  and $\alpha$ a
sensitivity parameter.  This parameter  is set to $-1/2$ on the basis
of a theoretical justification (improved bias  behavior) and  practical
experience (Abramson 1982).

The  final result being rather   insensitive  to the fine details   of
$f_p(x)$, it  would be natural  to adopt as a  pilot estimate the fixed
kernel estimate (eq.~\ref{eq_fixk})  with  optimal $h$  computed by
means of the normal reference rule  (eq.~\ref{eq_hopt}), and to compute
the final  estimate   by  applying  the adaptive   estimator   formula
(eq.~\ref{eq_lambda}).

However,  better estimations can be  obtained by  choosing the optimal
value  of  the   smoothing   parameter  $h$ by  means  of  data--based
algorithms.  Among   those proposed, unbiased cross--validation
(Rudemo 1982) and bootstrap cross validation  (Taylor 1989) are of
practical interest.  The former attempts to minimize the integrated
square error~:
\begin{eqnarray}
{\rm ISE}&=&\int [f(x)-\hf(x)]^{2}\dx \nonumber \\
&=&||\hf||_2+||f||_2-2\int f(x)\hf(x) \dx,
\end{eqnarray}
where  $||-||_2$   indicates the $L_2$  norm.    This is equivalent to
minimizing the quantity $||\hf||_2-2  E[\hf]$.  To obtain an estimate of
the   expected  value of  $\hf(x)$,    Rudemo considered the functions
$\hf_{-i}(x)$  obtained as  estimates on $n-1$  points   in the sample
excluding $x_i$. Hence, he proposed to minimize the quantity~:
\begin{equation}
{\rm UCV}(h)=||\hf||_2-\frac2n \sum_{i=1}^{n} \hf_{-i}(x_i).
\end{equation}
As    regards     Taylor's   (1989)     approach,   random  samples
$\{x_1^*,x_2^*,\ldots,x_n^*\}$ are   drawn  from the  candidate kernel
density estimate $\hf(x)$ (``smoothed bootstrap samples'').  Then, the
quantity~:
$$
E[\hf_*(x)-\hf(x)]=E[\frac1n \sum
K_h(x-x_i^*)-\frac1n \sum K_h(x-x_i)]^2 
$$
is computed, where  $\hf_*(x)$ is the estimate  on  the random sample.
If these samples come from the  empirical density (bootstrap samples),
this quantity evaluates only the  variance  of the estimate. The  bias
introduced by this ``smoothed bootstrap'' mimics the true unknown bias
related to the chosen smoothing parameter $h$.

We decided to  adopt  the UCV  algorithm for  reasons of computational
efficiency, and also taking into account the difficulty of obtaining one single
minimum value of $h$ with  other cross--validation methods.  The first
use of the UCV algorithm was made by Pisani (1993).

\section{Maximum Penalized Likelihood estimator.}

Maximizing the quantity $\sum f(x_i) - \alpha  \int (f''')^2$ with the
constraint $\int  \exp    (f(x)) \dx   =1$  can  be   treated  as   an
unconstrained   maximization  of   the    strictly  concave   function
(Silvermann 1986)~:
\begin{equation}
\sum f(x_i) - \alpha \int (f''')^2 - N \int \exp(f)
\label{unc_eq}
\end{equation}
It  is   possible to  avoid some of     the numerical and mathematical
difficulties of the MPL estimators by  replacing the integrals of this
equation with  approximations  on a  finite  interval $[a,b]$  (Scott,
Tapia  \&  Thompson 1980).  Thus, one  can  set  $f(a)=f(b)=0$  if the
interval is somewhat larger than the  range of all the observations or
one can mirror the data.

A discrete representation of  (\ref{unc_eq}) on a  uniform grid of $m$
evenly   spaced points with   corresponding  values  denoted by  $f_j$
($j=1,m$) is~:
$$ 
\sum_{i=1}^N f(x_i) -
\frac{\alpha} {\delta^5} \sum_{j=2}^{m-2} (-f_{j-1}+3 f_j -3 f_{j+1} +
f_{j+2})^2 - N \sum_{j=1}^m \epsilon_j f_j 
$$ 
with  $\delta = (b-a)/m$ and   $\epsilon_j=\delta$ for each $j$ except
for $\epsilon_1=\epsilon_m=\delta/2$. In the first term, $f(x_i)$ is a
linear approximation  between the points  of  the grid  which contain
$x_i$.  Starting with a uniform guess function,  one can maximize this
expression by varying the values  of the parameters  $f_j$.  As in the
case of the adaptive  kernel, we  can  choose an   optimal value  of  the
smoothing  parameter  with a data-based algorithm.   For instance,  the
unbiased cross validation  estimate of  $\alpha$   is the value   that
minimizes the function~:
\begin{equation}
{\rm UCV}(\alpha)=\int \hf^2(x) \dx - \frac2N \sum_{i=1}^N \hf_{-i}(x_i),
\end{equation}
where $\hf_{-i}$ is an estimate of $f$ constructed by leaving out the
single datum $x_i$.

\begin{figure*}
%\picplace {5.5cm}
\centerline{\psfig{file=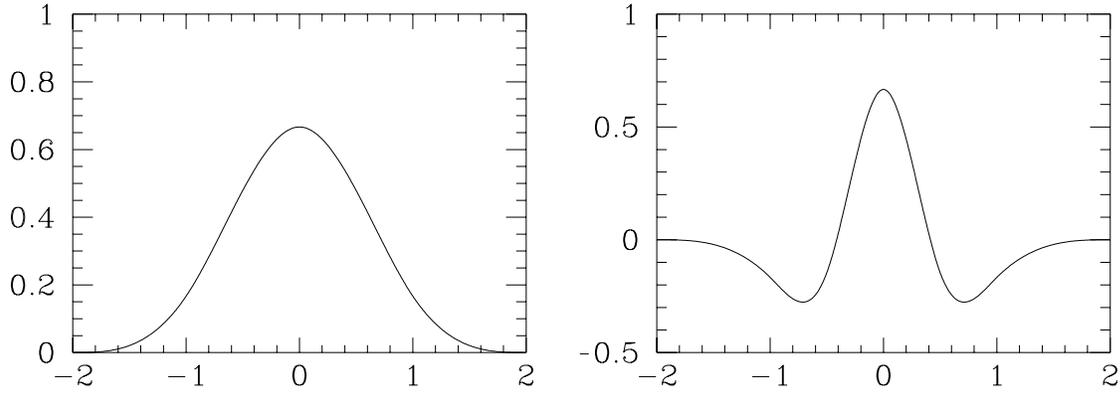,height=5.5cm}}
\caption []{The cubic B spline scaling function (left) and the related 
mother wavelet (right).}
\end{figure*}

\section{Wavelet estimator}

Among the  several algorithms which  are  available for computing  the
wavelet transform of a  one-dimensional function, the so-called  ``\`a
trous'' algorithm   makes  use of   undecimated data.   Although  this
intrinsic redundancy disqualifies it  for data compression purposes,   no
aliasing occurs in case of thresholding in  the wavelet space and this
algorithm  therefore  appears    very  well    suited for applications
requiring invariance  under  translations.  We thus decided  to
apply this algorithm.   \\ It is based  on a dyadic scheme ($a_i=2^i$)
where each approximation is computed   from the previous one with  finer
details, using the same low-pass discrete  filter $\{h(n)\}$.  Embedded
interpolations are required to do so in a  rigorous way, which implies
that  the function  $\phi(x)$   must  satisfy the following   dilation
equation~:
\begin{equation}
\frac12\ \phi(\frac x2) = \sum_n\ h(n)\ \phi(x-n).
\end{equation}
This is the case for a cubic B-spline~: 
$$ B_3(x)= \frac{1}{12}\ (\ |x-2|^3-4|x-1|^3+6|x|^3-4|x+1|^3+|x+2|^3\ ),$$
which also has  interesting  additional properties.  First, its compact
support provides  a local description  of the data.  Then, its regular
and symmetric shape with  a single bump  leads to a  wavelet transform
with  at  most two  small spurious  and  misleading negative secondary
peaks in the case of bright features (cf.  Figure~9).  This is very useful
for unambiguous vision and detection purposes.  Finally, this function
looks  like  a  Gaussian,  resembling the  features  which are  usually
sought for in   astronomical signals, and  it makes possible an isotropic
two-dimensional analysis from a tensorial product of spaces.  Thus, we
decided to adopt $B_3(x)$ as the  scaling function.  \\ From the definition  of
$\hf_{a_i}(k)$, and making use of the dilation equation, the successive
set of  approximations  $\hf_{a_{i+1}}$ can be computed  by convolving
the function $\hf_{a_i}$ with the filter $H_i$ according to the formula~:
\begin{equation}
\hf_{a_{i+1}}(k) = \sum_n\ h(n)\ \hf_{a_i}(k+2^i n) = 
(H_i\circ\hf_{a_i})\ (k),
\end{equation} 
where  $h_0=3/8$,  $h_{-1}=h_1=1/4$   and  $h_{-2}=h_2=1/16$ for   the
$B_3(x)$ function.   Similarly, relation  (\ref{coeff_eq}) can be
rewritten as~:
\begin{eqnarray}
W_{a_i}(k)& =& \hf_{a_i}(k) - \sum_n\ h(n)\ \hf_{a_i}(k+2^i n)\nonumber \\
          & =& \sum_n\ g(n)\ \hf_{a_i}(k+2^i n) = (G_i\circ\hf_{a_i})\ (k).
\end{eqnarray}
Hence,  one can write  the direct  relation $W_{a_i}(k)=T_i[\hf_0(k)]$
between the wavelet coefficients $W_{a_i}(k)$  and the discrete signal
$\hf_0(k)$ by defining an operator $T_i$ such as~:
\begin{equation}
 T_i=G_i \circ H_i \circ \cdots \circ H_1. 
\label{T0_eq}
\end{equation}

Once    significant   wavelet  coefficients    have    been   selected
(cf. discussion in \S~2.3), a  density estimate is computed by using  the
conjugate  gradient   technique, which looks  for  the  solution with the
minimal  energy  for  the  difference  between  initial   and restored
coefficients inside  significant   domains.   Basically, the  algorithm
consists in computing the estimate $\hf(x)$  by means of the iterative
relation~:
\begin{eqnarray}
\hf^{(0)}(k) & = & \widetilde{A}[\mathcal{W}\mathnormal{_t]} \nonumber \\
\hf^{(n)}(k) & = & \hf^{(n-1)}(k) + \alpha^{(n)} r^{(n)}(k),
\end{eqnarray}
where $\alpha^{(n)}$ is a convergence parameter and $r^{(n)}$ indicates
the residual signal at step $n$ defined as~:
\begin{equation}
r^{(n)}(k)=\widetilde{A}[ \mbox{$\mathcal W$}_t - A[ \hat{f}^{(n)}_{>0} ] ]
+\beta^{(n)} r^{(n-1)}(k),
\label{eq_rest}
\end{equation}
with $\beta^{(n)}$ a second convergence parameter, set to zero for the
first iteration.  The operator $\widetilde{A}$, which is equal to
\begin{equation}
\widetilde{A}[ \mathcal{W}\mathnormal ] = \sum_{i=1}^N\ 
(H_1 \circ \cdots \circ H_i)\ W_{a_i},
\end{equation}
transforms a set of wavelet coefficients $\mathcal{W}$ into a function
in direct space.  It  is the adjoint operator of  $A = P \circ T$, the
composition of the projection  and of the wavelet  transform operators
(see  eq.~\ref{T0_eq}).\\ Negative values   may arise due  to negative
wavelet coefficients surrounding high peaks,  which is unlikely  since
the number density estimate  must be a  positive function.  So at each
step the solution $\hf^{(n)}(k)$ has to be thresholded in order to get
a positive  estimate  $\hf^{(n)}_{>0} (k)$.   However, such  a  strong
thresholding   may  lead  to   some  discontinuities at  zero-crossing
locations and it is  inoperative  when a constant  density  background
exists or for  removing positive peaks coming from  deep holes in  the
data.  To  overcome these intrinsic difficulties,  one  may rely on an
iterative  structure subtraction, as  further explained  in \S~3.5, or
perhaps on an asymptotic positivity  constraint. These options will be
tested in the next future.

\end{document}